\documentclass[a4paper,times, fleqn,5p]{cas-dc}

\usepackage[numbers,sort&compress]{natbib}
\usepackage{subfigure}
\usepackage{threeparttable}
\usepackage{rotating}
\usepackage[spanish,english]{babel}
\usepackage{optidef}
\usepackage{verbatim}
\usepackage{multirow}
\usepackage{enumerate}
\usepackage{booktabs}
\def\tsc#1{\csdef{#1}{\textsc{\lowercase{#1}}\xspace}}
\tsc{WGM}
\tsc{QE}
\tsc{EP}
\tsc{PMS}
\tsc{BEC}
\tsc{DE}
\begin{document}
	
	\let\WriteBookmarks\relax
	\shorttitle{Graph-Time Spectral Analysis for Atrial Fibrillations}
	\shortauthors{Miao~Sun et~al.}
	
	\title [mode = title]{Graph-Time Spectral Analysis for Atrial Fibrillation}

	\author[1]{Miao~Sun}
	\cormark[1]

	\ead{M.Sun@tudelft.nl}

	\address[1]{Faculty of Electrical Engineering, Mathematics and Computer Science, Delft University of Technology, The Netherlands.}
	
	\author[1]{Elvin~Isufi}
	
	\author[2]{Natasja~M.S.~de~Groot}

	\address[2]{Department of Cardiology, Erasmus University Medical Center, The Netherlands.}
		
	\author[1]{Richard~C.~Hendriks}

	\cortext[cor1]{Corresponding author}

	\begin{abstract}
		Atrial fibrillation  is a clinical arrhythmia with multifactorial mechanisms still unresolved. Time-frequency analysis of epicardial electrograms  has been investigated to study atrial fibrillation. However, deeper understanding of atrial fibrillation can be achieved if the spatial dimension can be incorporated. Unfortunately, the physical models describing the spatial relations of atrial fibrillation signals are complex and non-linear; hence, the conventional signal processing techniques to study electrograms in the joint space, time, and frequency domain are less suitable.
		In this study, we  wish to put forward a radically different approach to analyze atrial fibrillation with a higher-level model. This approach relies on graph signal processing to represent the spatial relations between epicardial electrograms and put forward a graph-time spectral analysis for atrial fibrillation. To capture the frequency content along both the time and graph domain, we proposed the joint graph and short-time Fourier transform. The latter allows  us to analyze the spatial variability of the electrogram temporal frequencies. With this technique,  we have found that the spatial variation of the atrial electrograms decreases during atrial fibrillation due to the reduction of the high temporal frequencies of the atrial waves.  The proposed analysis further confirms that the ventricular activity is smoother over the atrial area compared with the atrial activity. Besides using the proposed graph-time analysis to conduct a first study on atrial fibrillation, we applied it to the cancellation of ventricular activity from atrial electrograms.  Experimental results on simulated and real data further corroborate the findings in this atrial fibrillation study.
		
		%
		
	\end{abstract}

	\begin{keywords}
		Atrial fibrillation \sep Graph signal processing\sep Spectral analysis\sep Atrial activity extraction\sep Graph-time signal processing
	\end{keywords}

	\maketitle
	
	\section{Introduction}
	Atrial fibrillation  is a cardiac arrhythmia characterized by rapid and irregular atrial beating and is correlated with stroke and sudden death \cite{january20142014,miller2005cost,miyasaka2007mortality}. Yet, the mechanisms underlying atrial fibrillation  remain still unresolved and challenging to model.
	To analyze the disease, different signal processing methods have been applied to the  non-invasive  body surface electrocardiograms (ECGs), or to the invasive epicardial or  endocardial electrograms \cite{ nitta2004concurrent, barbaro2001mapping, calcagnini2006descriptors, barbaro2002measure, teuwen2018quantification}. The epicardial electrogram (EGM) is measured directly on the heart's surface through multiple electrodes and has a higher spatial resolution compared with ECGs. This improved resolution makes EGMs appealing to analyze atrial fibrillation  over both space (heart surface) and time. The methods proposed in the current work concern EGM data. 
	
	Although different studies have analyzed electrograms data in  time and frequency domain  \cite{everett2001frequency, jacquemet2006analysis,bollmann1998frequency,xi2004atrial, houben2010analysis}, there remain many open questions that require alternative and novel tools to investigate atrial fibrillation. Experience in signal processing suggests that  incorporating the spatial dimension into the time-frequency analysis may yield improved insights on the atrial activity. However, the physical models for spatial propagation are relatively complex and non-linear; hence, rendering conventional signal processing methods  less suitable for a  joint space, time, and frequency domain analysis \cite{brandstein2013microphone, van1988beamforming}. It is also difficult to use the physical models for extracting useful information, e.g., activation time or conductivity  \cite{lombardo2016comparison}. 
	
	In this work, we wish to suggest a novel approach to model epicardial electrograms at a higher abstraction level.
	This approach represents the spatial relation of different epicardial electrograms through a graph and relies on graph signal processing  to investigate electrograms in the joint space, time, and frequency domain. We conduct a first study with the proposed framework   to identify spectral differences between sinus rhythm (normal heart rhythm) and atrial fibrillation, and between atrial and ventricular activities. We also leveraged the proposed graph model to remove ventricular components from the raw EGM measurements.

	\textit{Graph-time signal processing:}
	Graphs are natural tools to model data living in high-dimensional and irregular domains \cite{newman2018networks}. Graph signal processing provides a harmonic analysis for signals residing on the vertices of the graph and has been applied to brain signal analysis, Alzheimer classification, and body motion \cite{huang2016graph, medaglia2018functional, huang2018graph,hu2016matched, isufi2018blind,behjat2015anatomically, guo2017deep}. 
	However, despite showing promise, graph signal processing is still unexplored for heart-related problems.
	The EGM signals considered in this work are (spatially) high-dimensional measurements taken from epicardial sites of the atria during open-heart surgery \cite{yaksh2015novel}. Graph signal processing poses then itself as a valid candidate to account for the underlying mechanisms for analyzing atrial fibrillation. The atrial activity during atrial fibrillation is a complicated process for which it is hard to find a good and  tractable mathematical model. Graph signal processing can tackle this issue by formulating a high-level model for the atrial activity; hence, taking a step further towards exploring the atrial fibrillation behavior. The use of graphs to understand atrial fibrillation has also been considered in \cite{sun2014preliminary}. This work explored the association between different atrial regions through basic graph theory (e.g.,  graph topology, density, average degree), yet left  unexplored the processing of the signals on top of this graph. In this work, instead, we investigate  EGMs  through graph signal processing.

	The predominant tool in graph signal processing is the graph Fourier transform; a generalization of the temporal Fourier transform that provides a frequency interpretation for graph data. Similar to the time domain, the graph frequency components characterize the signal variation, now, over the graph and have shown to be useful to study biological activities \cite{huang2016graph, medaglia2018functional, huang2018graph}. Since the EGM varies with time, it is insufficient to consider the graph Fourier transform alone since it  analyzes  the  spatial variability for a fixed time instant. 
	To account for the temporal variability and capture the interaction between space and time, we can consider the so-called product graphs \cite{sandryhaila2014big}. A conceptual simpler alternative is to apply the graph Fourier transform on the data after applying the temporal Fourier transform (which tends to decorrelate the time-domain data). Since the electrogram is non-stationary, we use a joint graph and short-time Fourier transform to investigate the spatial properties of the temporal frequency content in a short-time period. Compared with the product graph method, working on the joint graph-time domain is simpler, and the analysis can be done independently per temporal frequency.

	\textit{Spectral EGM analysis:}
	We apply the graph-time spectral analysis framework to characterize the spectral properties  of the EGMs in the graph and time domain. We first evaluate the spatial variation   of the EGMs at different temporal frequencies during sinus rhythm and atrial fibrillation.
	During atrial fibrillation  this analysis showed that the high temporal frequencies of the atrial activity are reduced, leading also to a decrease of the spatial variation.  We also oberved that the spatial variation of the atrial activity is  higher than
	the spatial variation of the ventricular activity. We then used this difference in behavior  to extract the atrial activity from the mixed EGM measurement.
	
	\textit{Atrial activity extraction:}
	Electrograms measured on the atrial sites are naturally corrupted by the ventricular activity. The capability of a method to extract the atrial activity is fundamental to promote it for  atrial fibrillation studies.
	A common technique to extract the atrial activity  is  template matching such as average beat subtraction \cite{slocum1985computer}. Other techniques have also been proposed, such as adaptive ventricular cancellation \cite{rieta2007comparative}, principal component analysis \cite{raine2005surface}, and independent component analysis \cite{rieta2004atrial}. In this work, we develop a more effective algorithm to extract the atrial activity based on graph signal variation.

	\textit{Contribution and organization:}
	Altogether, this paper puts forward a radically different approach to analyze the epicardial electrograms from a higher abstraction level.
	This approach relies on graph signal processing and reveals features of biological and engineering interest.
	It also shows promise to remove interference from the atrial electrogram. More concretely, the contributions of this paper are: (i) To propose a high-level graph signal processing model for analyzing the epicardial electrogram data; (ii) To evaluate the temporal and spatial variation of  epicardial electrograms using a graph-time spectral analysis framework. This helps to: (ii-a) recognize  atrial fibrillation impact on the atrial activity; (ii-b) identify  differences between the atrial and ventricular activities; (iii) To propose a novel and effective atrial activity extraction algorithm based on the variations of the atrial  and ventricular activities over the graph.
	
	The rest of this paper is organized as follows. In Section II, we describe the data used in this work. In Section III, we  introduce the basic notation of graph signal processing and  the joint graph and short-time Fourier transform. In Section IV, we perform the graph-time spectral analysis on the electrograms under sinus rhythm and atrial fibrillation. In Section V, we present the atrial activity extraction algorithm and evaluate its performance on synthetic and real data. We discuss the paper contributions and future directions in Section VI and draw the conclusions in Section VII.
	
	\begin{figure}
		
		\centering
		\includegraphics[height = 5.5cm]{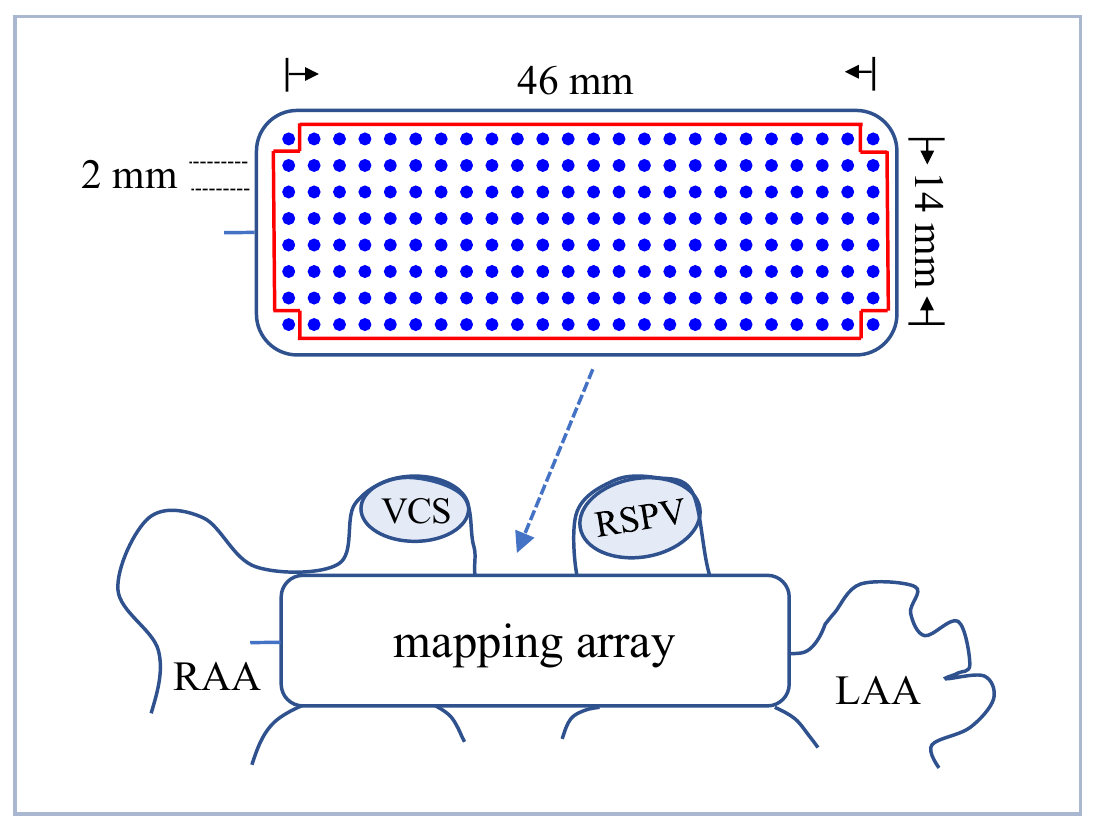} %
		\caption{The mapping array and the placement on the Bachmann's bundle area of the atria \cite{yaksh2015novel}. RAA: right atrial appendage; LAA: left atrial appendage; VCS: vena cava superior; RSPV: right superior pulmonary vein.}
		\label{fig:array}
	\end{figure}
	
	\section{Database}
	
	We used the epicardial electrogram data  measured on human atria during open-heart surgery as reported in \cite{yaksh2015novel}. Ten patients (aged 64$\pm$16; 20\% female) are analyzed in this study. Three patients underwent surgery due to aorta ascendens dilatation  and the remaining seven due to aortic valve and coronary artery disease; all patients did not have a reported  history of atrial fibrillation. The atrial fibrillation was  induced manually by rapid pacing in the right atrial free wall with the procedure detailed  in the original   publication \cite{yaksh2015novel}. We  remark that induced atrial fibrillation has also been used to investigate  the disease in \cite{sun2014preliminary} and \cite{everett2001frequency}.  For each patient, both sinus rhythm and atrial fibrillation data are recorded.
	
	Previous research has suggested that the Bachmann's bundle area is related to the pathophysiology of atrial fibrillation \cite{van2013bachmann}. However, this area is still one of the less understood area. Because of the connection with atrial fibrillation and the interesting research aspects,  we will hereinafter focus on the EGMs measured on this area.
	
	A mapping array of 8$\times$24 electrodes with an inter-electrode distance of 2 mm is used to collect data. 
	During the measurement phase, 188 electrodes record the EGMs; these are the electrodes in the red box in Figure 1. Three of the remaining electrodes are  used  to record the body surface ECG signal, the reference signal, and the calibration signal, respectively; the last electrode is not used. The electrogram  comprises five seconds of recordings during sinus rhythm and ten seconds during atrial fibrillation with a sampling rate of 1 kHz. 
	All measurements were taken in the Erasmus Medical Center, the Netherlands, during  the period 2014-2016 with procedures approved by the Medical Ethical Committee (MEC 2010-054 \& MEC 2014-393) \cite{van2016quest, lanters2015halt}. Further details about the data acquisition system are reported in \cite{yaksh2015novel}.
	
	\begin{figure}[]%
		\centering
		\subfigure[]{%
			\includegraphics[height = 6.5cm]{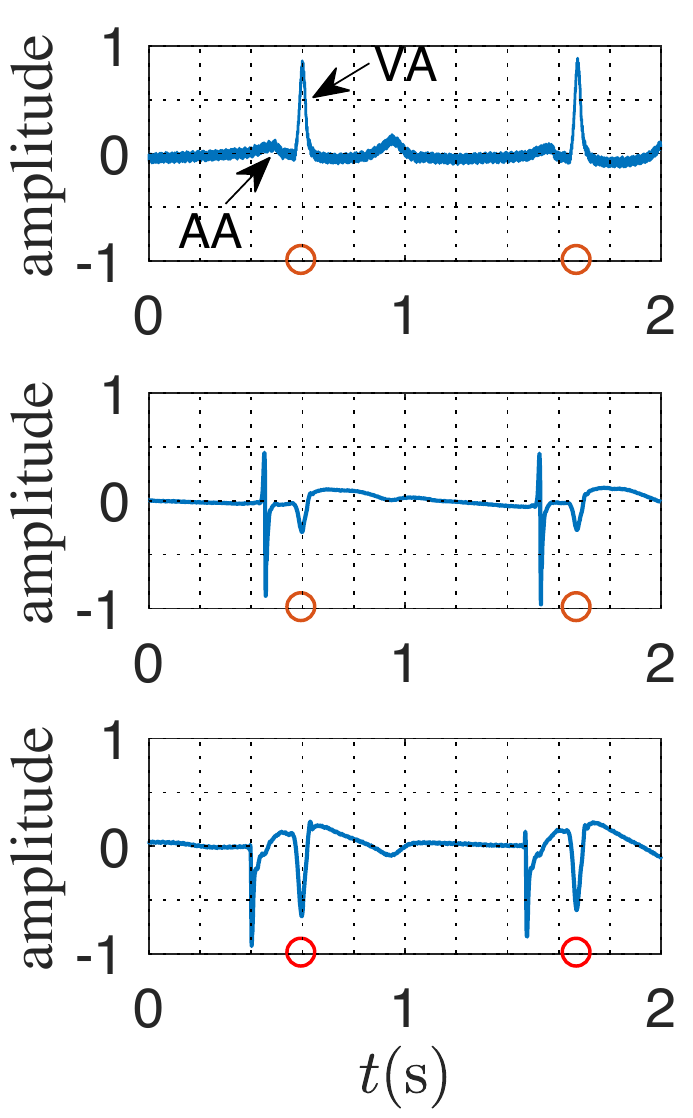}}%
		\qquad
		\subfigure[]{%
			\includegraphics[height=6.5cm]{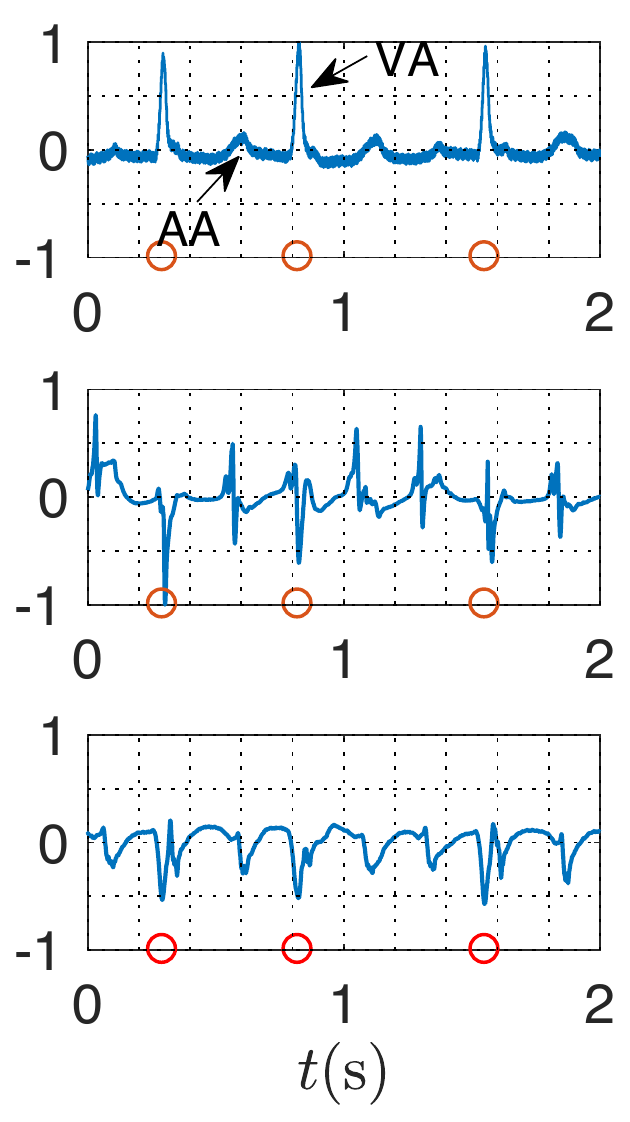}}%
		\caption{Examples of the body surface  electrocardiogram (ECG) and  epicardial electrogram (EGM) across time at one electrode during (a) sinus rhythm  and (b) atrial fibrillation. AA: atrial activity; VA: ventricular activity. \textit{Top}: ECG; \textit{middle} and \textit{bottom}: EGMs at different electrodes. The red circles mark the peak of the ventricular activity determined by the ECG measurements. }
		\label{fig:data}
	\end{figure}

	Figure \ref{fig:data} illustrates the ECGs and the EGMs measured on Bachmann's bundle  during sinus rhythm and atrial fibrillation for one patient.   In the ECG (top plots in Figures \ref{fig:data}(a) and \ref{fig:data}(b)), the high peaks indicate  the ventricular activity, while the lower peaks before them indicate the atrial activity. The atrial activity appears  weak compared with the ventricular activity. In the EGM measurements (middle and bottom plots in Figures \ref{fig:data}(a) and \ref{fig:data}(b)), the atrial activity is more pronounced, albeit short in duration. This difference  is due to spatial averaging occurring when measuring the atrial signal on the body surface, compared with when measuring it on the epicardium.
	
	From Figure \ref{fig:data}(b), we see that the atrial and the ventricular activities during atrial fibrillation are difficult to distinguish since they appear irregular and overlap. In other words,
	the  ventricular activity  will affect the analysis on the atrial activity; hence, extracting the atrial activity from the measurement is critical for atrial fibrillation research.

	The EGMs measured by the different electrodes (middle and bottom plots in Figure 2) show a time delay when measuring the atrial activity in  different positions. However, they do not show any obvious time delay when measuring the ventricular activity. This is because the mapping array for data measurements is close to the atria and far from the ventricle. Also,  the amplitudes of the ventricular activity are different at the different electrodes due to the propagation attenuation of the signal.

	The above illustration highlights the limitations of the body surface ECG--the atrial activity in there is   weak   and gets easily corrupted by noise; hence,  rendering the time-frequency analysis not reliable.
	Although proposed invasive methods measured a stronger atrial activity, they used low-resolution mapping arrays for the measurements  and analyzed the data only in time or temporal frequency domains \cite{ nitta2004concurrent, barbaro2001mapping, calcagnini2006descriptors, barbaro2002measure}. Differently, we consider high-resolution epicardial measurements and analyze the data in the joint space, time, and frequency domain. 
	
	\section{Theory}
	In this section, we recall the basic concepts on graph signal processing and introduce the joint graph and short-time Fourier transform.
	
	\subsection{Graph signal processing}

	\textit{Graphs and graph signals:} Consider a network represented by an undirected graph $\mathcal{G} = \left( \mathcal{V}, \mathcal{E}, \mathbf{W} \right)$, where $\mathcal{V} = ( v_{1} , \cdots, v_{K} )$ is the set of $K$ vertices, $\mathcal{E}$ is the edge set, and $\mathbf{W}$ is the graph adjacency matrix with entries $\mathbf{W}(i,j)=W_{i,j}$. Here, $W_{i,j} \ge 0$ represents the edge weight connecting vertices $v_{i}$ and $v_{j}$ and  $W_{i,j} = 0$  indicates  no connection between  vertices. The neighbor set of vertex $v_i$ is denoted as $\mathcal{N}_i$.
	The graph Laplacian matrix is $\mathbf{L} = \mathbf{D} - \mathbf{W}$, where $\mathbf{D}$ is the diagonal degree matrix with $D_{i,i} = \sum_{j=1}^{K}W_{i,j}$.

	A graph signal is a set of values over the vertices, i.e., it is  a mapping from the vertex set to the set of real numbers, $y:\mathcal{V} \rightarrow \mathbb{R}$.  The epicardial electrograms recorded by all electrodes of the mapping array is an example of a graph signal. Let $y_{i}(t)$  be the signal of vertex $v_i$ at time $t$ for $i = 1, \ldots, K$ and $t = 0, \ldots , T-1$. The graph signal at time instant $t$ is compactly represented by the $K\times 1$ vector $\mathbf{y}(t)= [ y_{1}(t), y_{2}(t), \ldots , y_{K}(t)]^{T}$.
	
	The electrical activities recorded by the electrodes of the mapping array are related to each other and form an electrical network. We constructed a graph for the mapping array by considering each electrode as a vertex. There are two ways to build the edges in the graph: (i) based on the data structure, e.g., correlation; (ii) based on physical properties, e.g., distance.

	To compare the sinus rhythm signal with the atrial fibrillation signal, we consider a  fixed graph structure for both situations. With the illustration in Figure 3(a), the edges  are determined by the electrodes position; each vertex is connected with its eight nearest neighbors.
	This expresses that an electrode (vertex) has strong similarities with the surrounding electrodes. In other words, this graph is build with the prior knowledge that under healthy conditions neighboring vertices are expected to record a similar signal. 
	The edge weights are based on the distance between two connected vertices. This is  a common approach in graph signal processing  when there is little prior knowledge of the graph signal. The edge weight is 
	\begin{equation}
	W_{i,j} = \left(\frac{d_{i,j}}{\alpha}\right)^{-1}
	\end{equation}
	where $d_{i,j}$ is the distance between two connected vertices and $\alpha$ is a scaling parameter. It is chosen as the smallest distance between two vertices to normalize the largest weight to one.

	\textit{Graph Fourier transform and smoothness:}
	The graph Laplacian matrix is symmetric, positive semidefinite, and accepts  the eigenvalue decomposition
	\begin{equation} 
	\mathbf{L} = \mathbf{U} \mathbf{\Lambda}  \mathbf{U}^{H}\\
	\end{equation} 
	where $\mathbf{U} = [\mathbf{u}_0,\mathbf{u}_1, \cdots, \mathbf{u}_{K-1}]$ is the set of orthonormal eigenvectors, $\mathbf{\Lambda}$ is the diagonal matrix of eigenvalues, and $(\cdot)^H$ is the Hermitian operator. 
	The eigenvalues  are sorted in increasing order $0=\lambda_{0} < \lambda_{1} \leq \cdots \leq \lambda_{K-1}$.
	
	\begin{figure}[]%
		\hspace*{\fill}
		\subfigure[]{%
			\includegraphics[trim=0cm 0cm 0cm 0cm,clip=true,width=2.6cm,height=3.6cm]{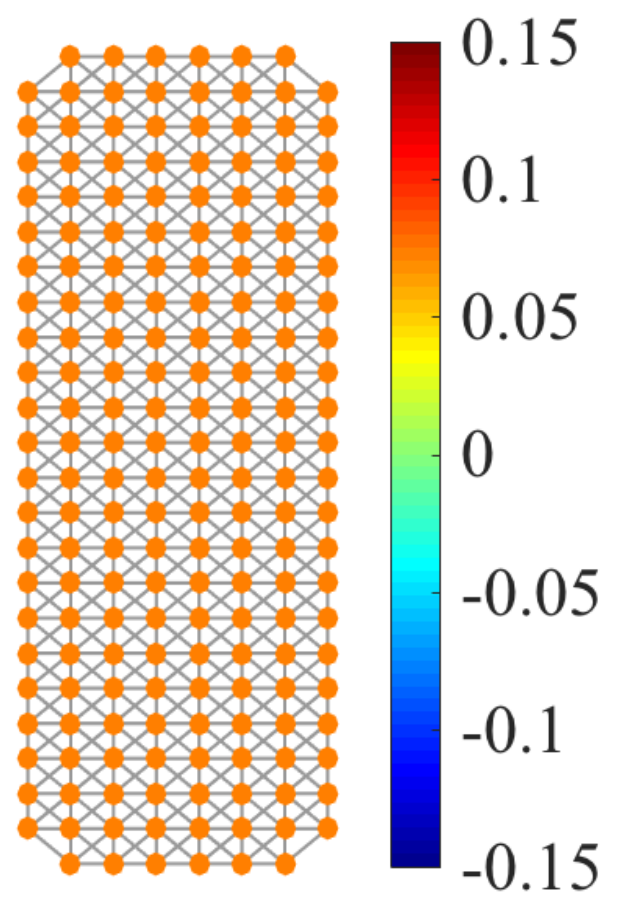}}%
		\hspace*{\fill}
		\subfigure[]{%
			\includegraphics[width=2.6cm,height=3.6cm]{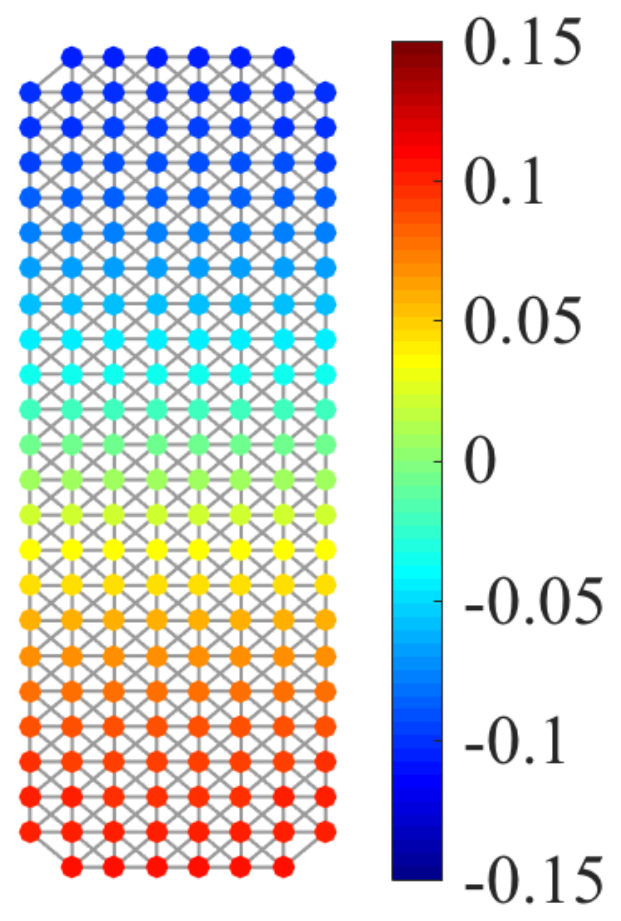}}%
		\hspace*{\fill}
		\subfigure[]{%
			\includegraphics[width=2.4cm,height=3.6cm]{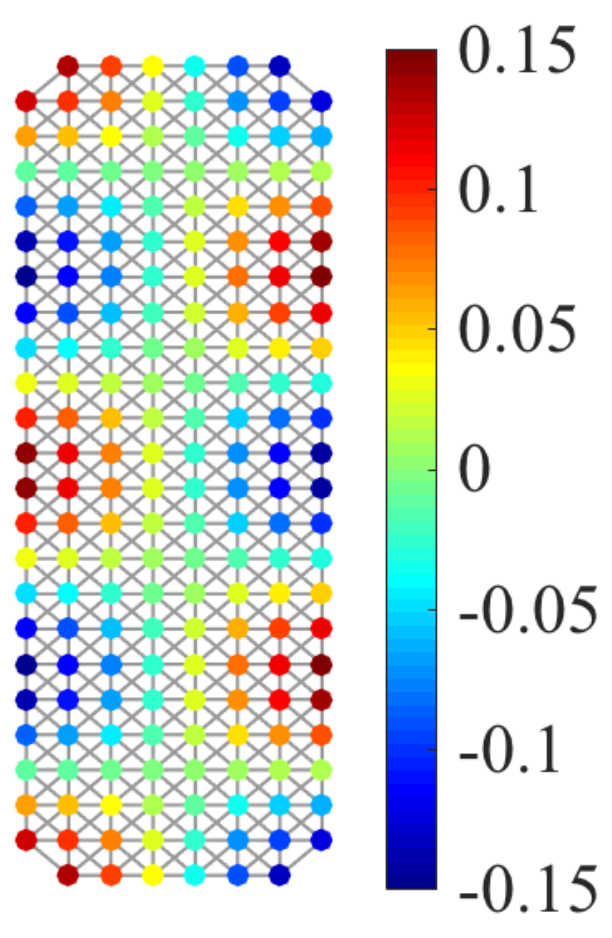}}%
		\caption{Different graph Laplacian eigenvectors of the graph. (a) $\mathbf{u}_0=1/\sqrt{K}\mathbf{1}$ is the constant eigenvector shown by the same color over all vertices; (b) $\mathbf{u}_1$ is a slow-varying eigenvector shown by a smooth transition from the top vertices to the bottom ones; (c) $\mathbf{u}_{9}$ is a faster-varying eigenvector over the graph  shown by the multiple variations in adjacent vertices.}
		\label{fig:eigenvectors}
	\end{figure}
	The graph Fourier transform (GFT) of the graph signal $\mathbf{y}(t)$ with respect to  Laplacian $ \mathbf{L} $ is 
	\begin{equation} 
	\widetilde{\mathbf{y}}(t) = \mathbf{U}^{H}\mathbf{y}(t)\\
	\end{equation} 
	where $\widetilde{\mathbf{y}}(t) =[\widetilde{y}(0,t), \widetilde{y}(1,t), \ldots, \widetilde{y}(K-1,t)]^H$ contains the GFT coefficients   $\tilde{y}(k,t)$ for $k = 0,...,K-1$. The inverse GFT is 
	\begin{equation} 
	\mathbf{y}(t) = \mathbf{U}\widetilde{\mathbf{y}}(t). \\
	\end{equation} 
	The GFT is a generalization of the temporal Fourier transform: for the graph being a cycle that represents the temporal axis of a periodic signal, the GFT matches the discrete Fourier transform \cite{sandryhaila2014big}. 
	In general, the GFT analyzes the signal variation over the graph for a fixed time instant. Since the transform (eigenvector) matrix $\mathbf{U}$ depends on the graph structure, it gives a harmonic decomposition for signals living in irregular domains where the traditional discrete Fourier transform cannot be applied.  For readers familiar with spectral network theory, the GFT can also be seen as the signal projection onto the Laplacian eigenspace.
	
	The GFT coefficients
	$\widetilde{\mathbf{y}}(t)$ for lower values of $k$  indicate how much the slower varying eigenvectors over the graph contribute to $\mathbf{y}(t)$. For larger  values of $k$, they indicate how much the  faster varying eigenvectors over the graph contribute to $\mathbf{y}(t)$. The coefficient $\widetilde{y}(0,t)$ indicates the contribution of the constant component (which is equal to $1/\sqrt{K}$ at each vertex) on $\mathbf{y}(t)$ \cite{shuman2013emerging}. Therefore, the index $k$ is also called  as the graph frequency index. Figure \ref{fig:eigenvectors} depicts three  eigenvectors of the considered graph: the eigenvector $\mathbf{u}_k$ changes more rapidly over adjacent vertices for larger $k$.

	Just like temporal bandlimited signals, we can define bandlimited graph signals. In many practical cases, the coefficients $\widetilde{y}(k,t)$ have only a few non-zero entries. A bandlimited graph signal $\mathbf{y}(t)$ is therefore defined as a graph signal with GFT coefficients \cite{shuman2013emerging}
	\begin{equation} 
	\widetilde{y}(k,t) = 0, \text{ for  } k>K_0\in \{0, \cdots, K-1\}
	\end{equation}
	implying  the signal has no content outside the graph frequency band of $\{0, K_0\}$.

	To measure the signal variation over the graph, the graph Laplacian quadratic form of $\mathbf{y}(t)$ is defined as \cite{shuman2013emerging}
	
	\begin{equation}
	\begin{split} 
	\text{V}_{\mathcal{G}}\left( \mathbf{y}(t)\right) &= \mathbf{y}(t) ^H \mathbf{L} \mathbf{y}(t)\\&=\sum_{i \in \mathcal{V}}\sum_{ j \in \mathcal{N}_i }{W_{i,j}\left(y(i,t)-y(j,t)\right)^2}.
	\label{eq:tv}
	\end{split}
	\end{equation}
	This quadratic form shows that the signal variation $\mathbf{y}(t)$ over the vertices for a fixed $t$ is a weighted sum of the difference between any two connected vertices. The edge weight indicates the contribution of a specific connection to the overall variation. If $\text{V}_{\mathcal{G}}\left( \mathbf{y}(t)\right)$ is small, it means the signal is smooth, i.e., it has similar values in adjacent vertices. If $\text{V}_{\mathcal{G}}$ is large, it means the signal  changes faster over the graph, i.e., it has different values in adjacent vertices. For the three eigenvector signals in Figure 3, we have that $0 = \text{V}_{\mathcal{G}}(\mathbf{u}_0) < \text{V}_{\mathcal{G}}(\mathbf{u}_1) < \text{V}_{\mathcal{G}}(\mathbf{u}_{9})$.

	\subsection{Joint STFT and GFT}
	The discussed graph signal processing framework considers only a single time instant and does not capture the correlation of the signal across time. Since the signals we study are time-varying and non-stationary, the joint graph and short-time Fourier transform is defined next to exploit the signal dependencies across both graph and time.
	In simple words, the short-time Fourier transform (STFT) is  applied first to transform the signal per vertex to the temporal frequency domain. This approximately decorrelates the data per vertex. Subsequently, the GFT is applied on the each  temporal frequency independently.

	Let us split the signal into $M$ temporal frames of length $T_M$ and let  $\mathbf{y}(\tau, t) \in \mathbb{R}^{K\times 1}$ be the graph signal in frame $\tau \in \{0, \ldots, M-1\}$ at time instant $t$, i.e., the signal of all electrodes at one time instant.
	We represent all signals recorded in frame $\tau$ through the compact matrix form
	\begin{equation}
	\begin{split}
	{\bf{Y}}(\tau) = [\mathbf{y}(\tau, {\tau}T_M),&\mathbf{y}( \tau, {\tau}T_M + 1),\cdots,\\&\mathbf{y}(\tau, (\tau+1)T_M-1)] \in \mathbb{R}^{K \times T_M}
	\label{Eq:time_sample_matrix}
	\end{split}
	\end{equation}
	where the $i$th row of ${\bf{Y}}(\tau)$ corresponds to the time-varying signal measured by the $i$th electrode in frame $\tau$. 
	
	For the STFT transform, we consider $F$ temporal frequency bins and  apply a temporal window followed the discrete temporal Fourier transform to each row of ${\bf{Y}}(\tau)$. The STFT coefficient matrix of (7) at frame $\tau$ is 
	\begin{equation}
	\hat{\mathbf{Y}}(\tau)  = [\mathbf{\hat{y}}(\tau,0 ), \mathbf{\hat{y}}\left(\tau, 1 \right), \ldots, \mathbf{\hat{y}}\left(\tau, F-1 \right)]\in \mathbb{C}^{K\times F}
	\end{equation} 
	The $f$th column of ${\hat{\bf{Y}}}(\tau)$ with  $f\in\{0,...,F-1\}$ is given by
	\begin{equation}
	\mathbf{\hat{y}}\left(\tau, f \right) = \left[ \hat{Y}_{1} \left( \tau, f \right), \hat{Y}_{2} \left( \tau, f \right), \cdots, \hat{Y}_{K} \left( \tau, f \right) \right]^{H} \in \mathbb{C}^{K},
	\end{equation}
	which represents the temporal frequency components of all vertices in frame $\tau$ and frequency bin $f$. The GFT is then applied to each column of $\hat{\mathbf{Y}}(\tau)$ separately to achieve the joint STFT and GFT matrix
	\begin{equation}
	\widetilde{\mathbf{Y}}\left( \tau\right)  = \mathbf{U}^{H} \mathbf{\hat{Y}}\left(\tau\right)
	\end{equation}
	with $\widetilde{\mathbf{Y}}(\tau) = [\widetilde{\mathbf{y}}\left(\tau, 0 \right), \widetilde{\mathbf{y}}\left(\tau, 1 \right), \ldots, \widetilde{\mathbf{y}}\left( \tau, F-1 \right)]$. The $f$th column of $\widetilde{\mathbf{Y}}(\tau)$, i.e.,  $\widetilde{\mathbf{y}}\left(\tau, f \right)$,  is the GFT of $\mathbf{\hat{y}}\left(\tau, f \right)$;
	the $k$th element $\widetilde{y}\left(k, \tau, f \right)$  corresponds to the graph frequency index $k$. For a low value of $k$, this coefficient  indicates how much the slowly varying component over the graph contributes to the temporal frequency component $f$ in time frame $\tau$. Therefore, the joint   coefficient quantifies  the variation over the graph of a   temporal frequency   in a short-time period.  In other words, each coefficient indicates the EGM variation over space and time. These values  will be different when analyzed, for instance, during sinus rhythm compared with atrial fibrillation and they will  reveal patterns of space-time variability of the disease.
	
	To obtain again  the time-vertex signal $\mathbf{Y}(\tau)$ [cf. (\ref{Eq:time_sample_matrix})] from the joint transform representations, we first apply the inverse GFT  to $\widetilde{\mathbf{Y}}\left(\tau\right)$ as
	\begin{equation}
	\mathbf{\hat{Y}}\left(\tau\right) = \mathbf{U}\widetilde{\mathbf{Y}}\left( \tau\right)
	\end{equation}
	and get the STFT  matrix $\hat{\mathbf{Y}}(\tau)$. Then, we apply the inverse STFT and overlap-adding to reconstruct the entire time domain signal from the segmented frames.
	
	Similarly to  (\ref{eq:tv}), the variation of the temporal frequency components $\hat{\mathbf{y}}\left(\tau, f \right)$ over the graph can be quantified by the Laplacian quadratic form
	\begin{equation}
	\begin{split}
	\text{V}_{\mathcal{G}}(\hat{\mathbf{y}}\left(\tau, f \right))& =  \hat{\mathbf{y}}\left(\tau, f \right)^{H}\mathbf{L}\hat{\mathbf{y}}\left(\tau, f \right)\\ & = \sum_{i \in \mathcal{V}}\sum_{j \in \mathcal{N}_i }{W_{i,j}(\hat{Y}_{i}(\tau,f)-\hat{Y}_{j}(\tau,f))^2}.
	\end{split}
	\label{eq:tv_joint}
	\end{equation}
	The measure in (\ref{eq:tv_joint}) quantifies the variation over the graph of each temporal frequency $f$ in the time frame $\tau$. Since the variation can differ in different temporal frequencies, we consider the normalized variation
	\begin{equation}
	\text{V}_{{\mathcal{G}},{\text{n}}}(\hat{\mathbf{y}}\left( \tau, f \right)) = \frac{{ \hat{\mathbf{y}}}\left( \tau, f \right)^{H}{\mathbf{L}}{ \hat{\mathbf{y}}}\left( \tau, f \right)}{{ \hat{\mathbf{y}}}\left( \tau, f \right)^{H}{ \hat{\mathbf{y}}}\left( \tau, f \right)}= \frac{{ \widetilde{\mathbf{y}}}\left( \tau, f \right)^{H}{\mathbf{\Lambda}}{ \widetilde{\mathbf{y}}}\left( \tau, f \right)}{{ \widetilde{\mathbf{y}}}\left( \tau, f \right)^{H}{ \widetilde{\mathbf{y}}}\left( \tau, f \right)}
	\label{eq:tvn}
	\end{equation}
	where the second equality holds from the GFT.
	
	We will in the sequel use this joint transform to analyze the  EGMs  in three domains: the time domain, the temporal frequency domain, and the graph frequency domain.

	\section{Graph-time spectral analysis}
	In this section, we perform a spectral analysis on the EGMs during both sinus rhythm and atrial fibrillation. We first conduct a separate short-time Fourier transform  and graph Fourier transform analysis. Then, we conduct a joint transform analysis.

	\subsection{STFT analysis}
	For the STFT, we used a Hanning window of 0.1 s  with 50\% overlap. The window size  depends in general on the information we need to extract. For our analysis, we set the  length  equal to the approximate duration of the atrial  and  ventricular activities; both having a duration around 0.1 s.
	We analyzed the signal energy distribution over both time and temporal frequencies through the normalized energy 
	\begin{equation} 
	E(\tau, f)  = 10 \text{log}_{10}(|\hat{Y}_{\text{n}}(\tau, f)|^2) = 10 \text{log}_{10}\left(\frac{{|\hat{Y}(\tau,f)|^2 }}{{|\hat{Y}(\tau,f)|_{\text{max}}^2 }} \right)
	\label{eq:nor_amplitude}
	\end{equation} 
	where $\hat{Y}(\tau,f)$ is the STFT coefficient, $|\hat{Y}(\tau,f)|_{\text{max}}$ is the maximum  amplitude, and $|\hat{Y}_{\text{n}}(\tau, f)|$ is the normalized signal amplitude.

\begin{figure}%
		\centering
		\subfigure[]{%
			\includegraphics[width=4cm,height=2.8cm]{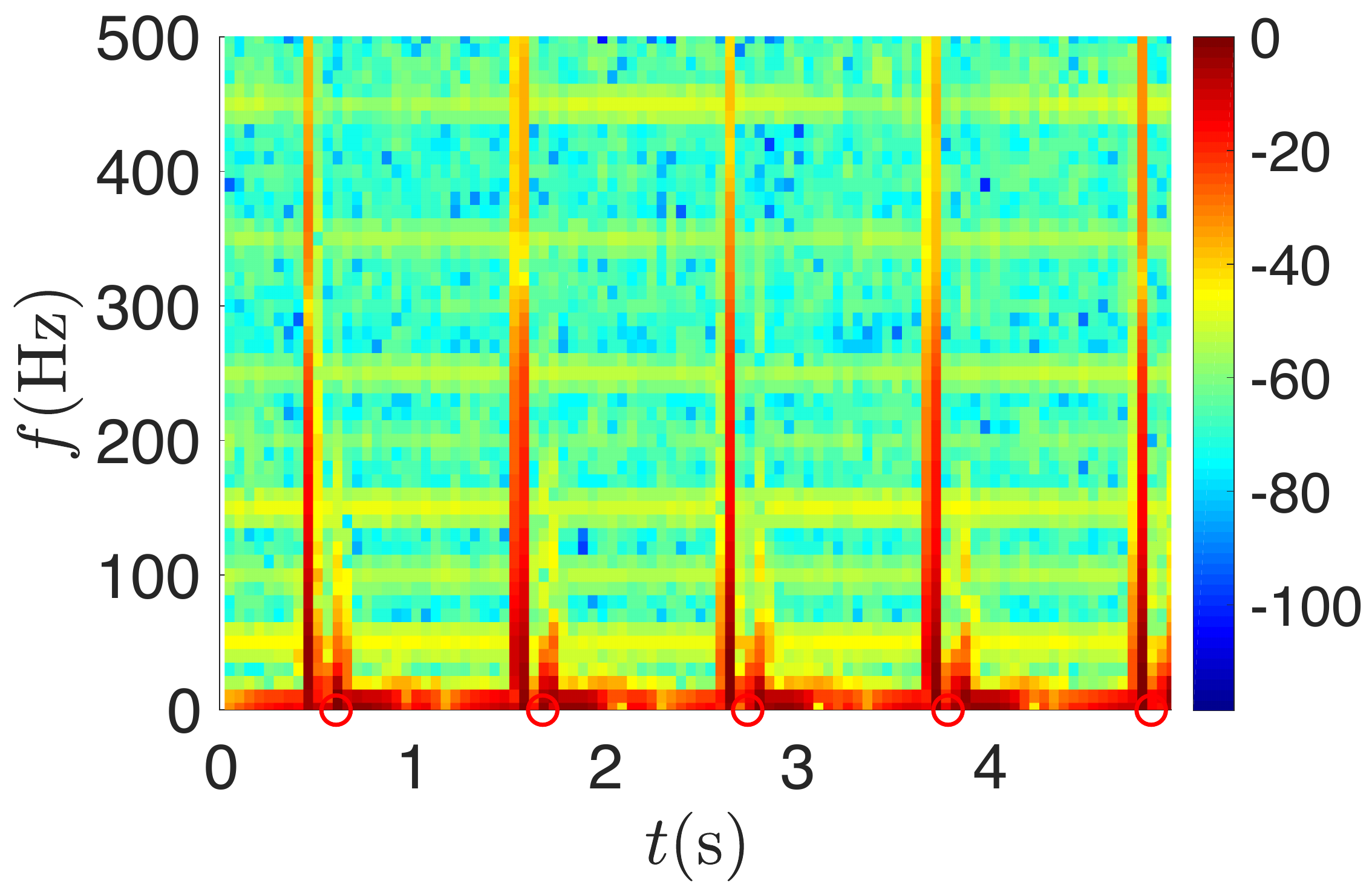}}
		\hspace*{\fill}
		\subfigure[]{%
			\includegraphics[width=4cm,height=2.8cm]{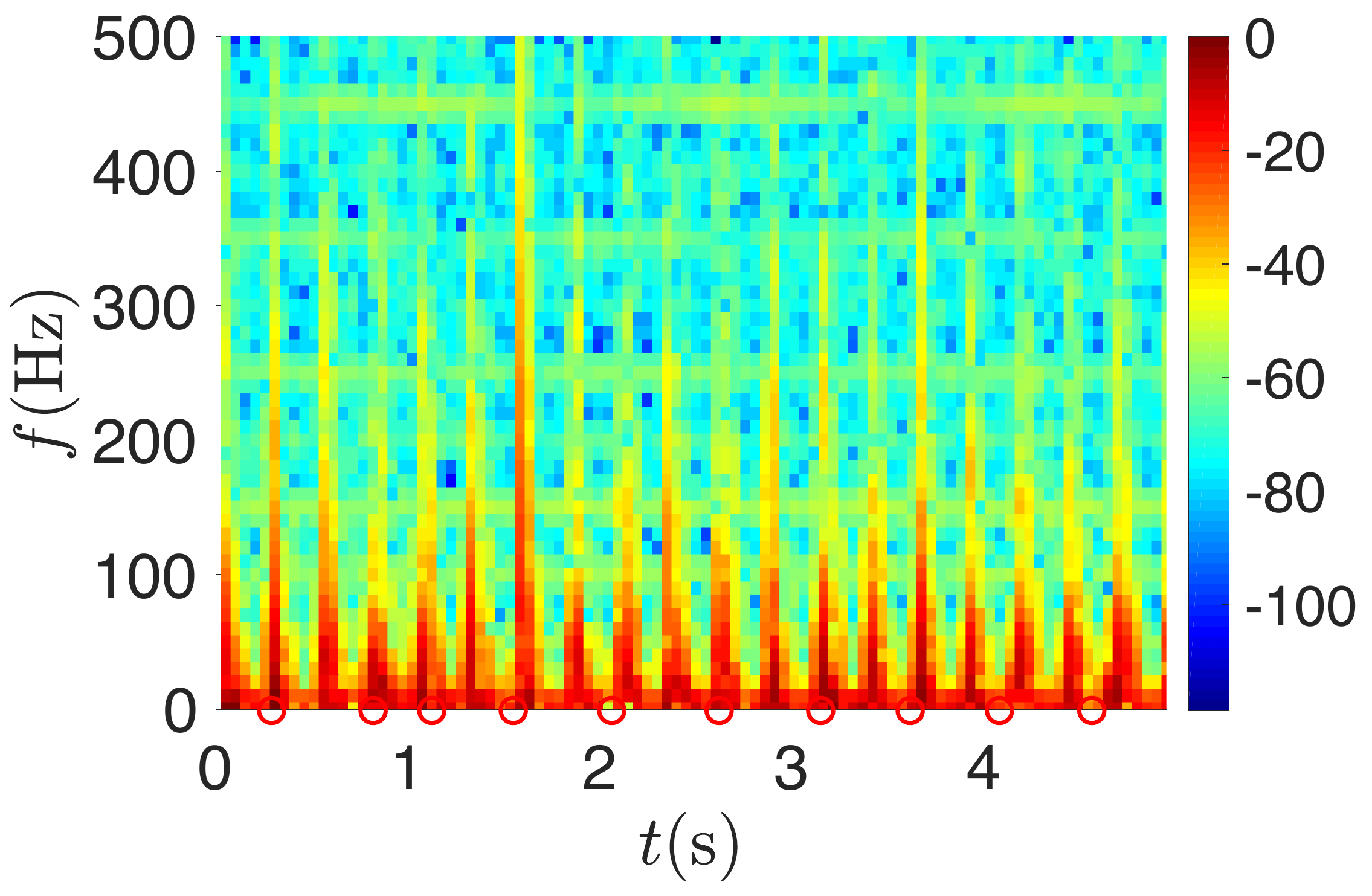}}
		\qquad
		\subfigure[]{%
			\includegraphics[width=4cm,height=2.5cm]{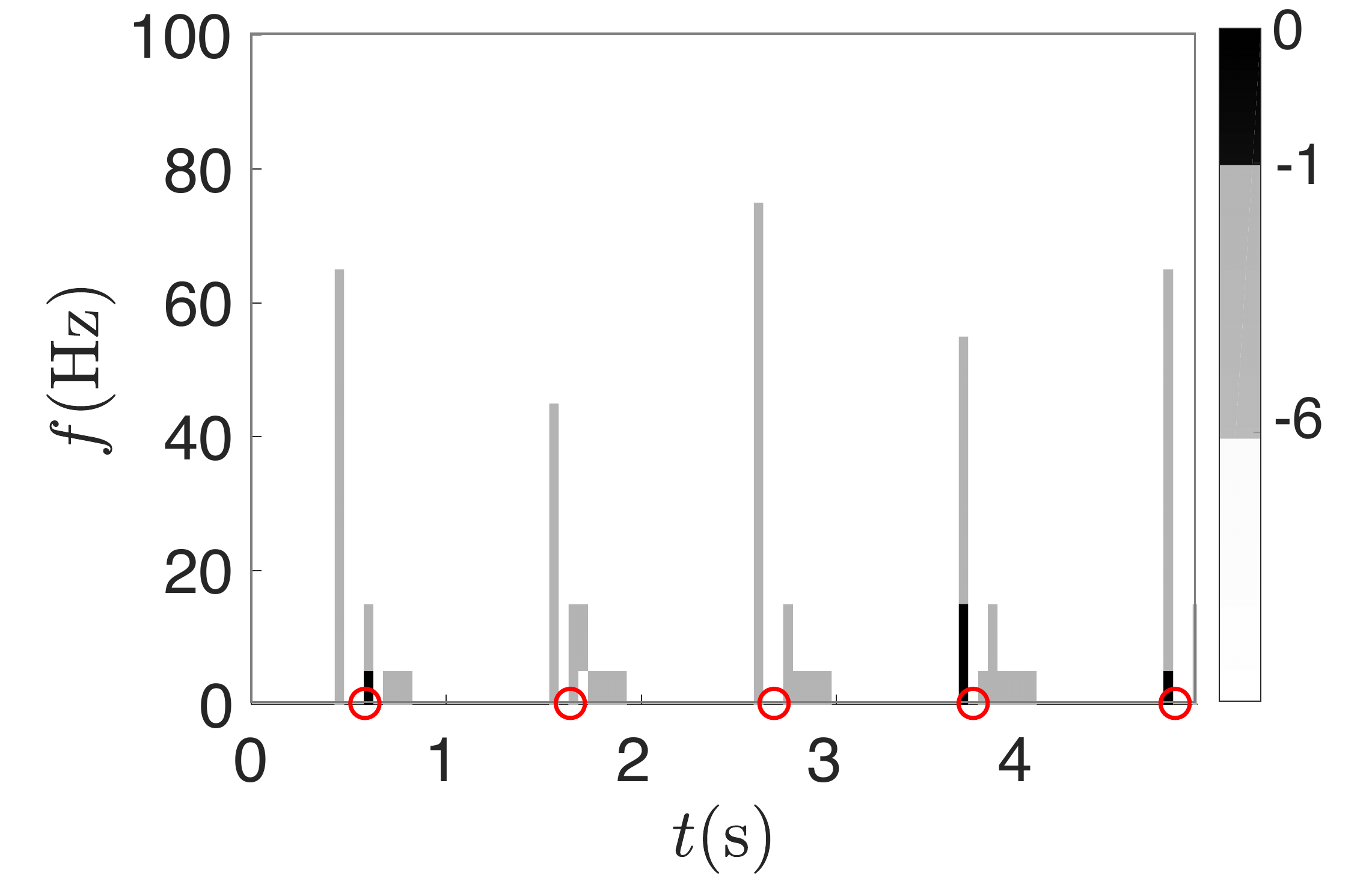}}
		\hspace*{\fill}
		\subfigure[]{%
			\includegraphics[width=4cm,height=2.5cm]{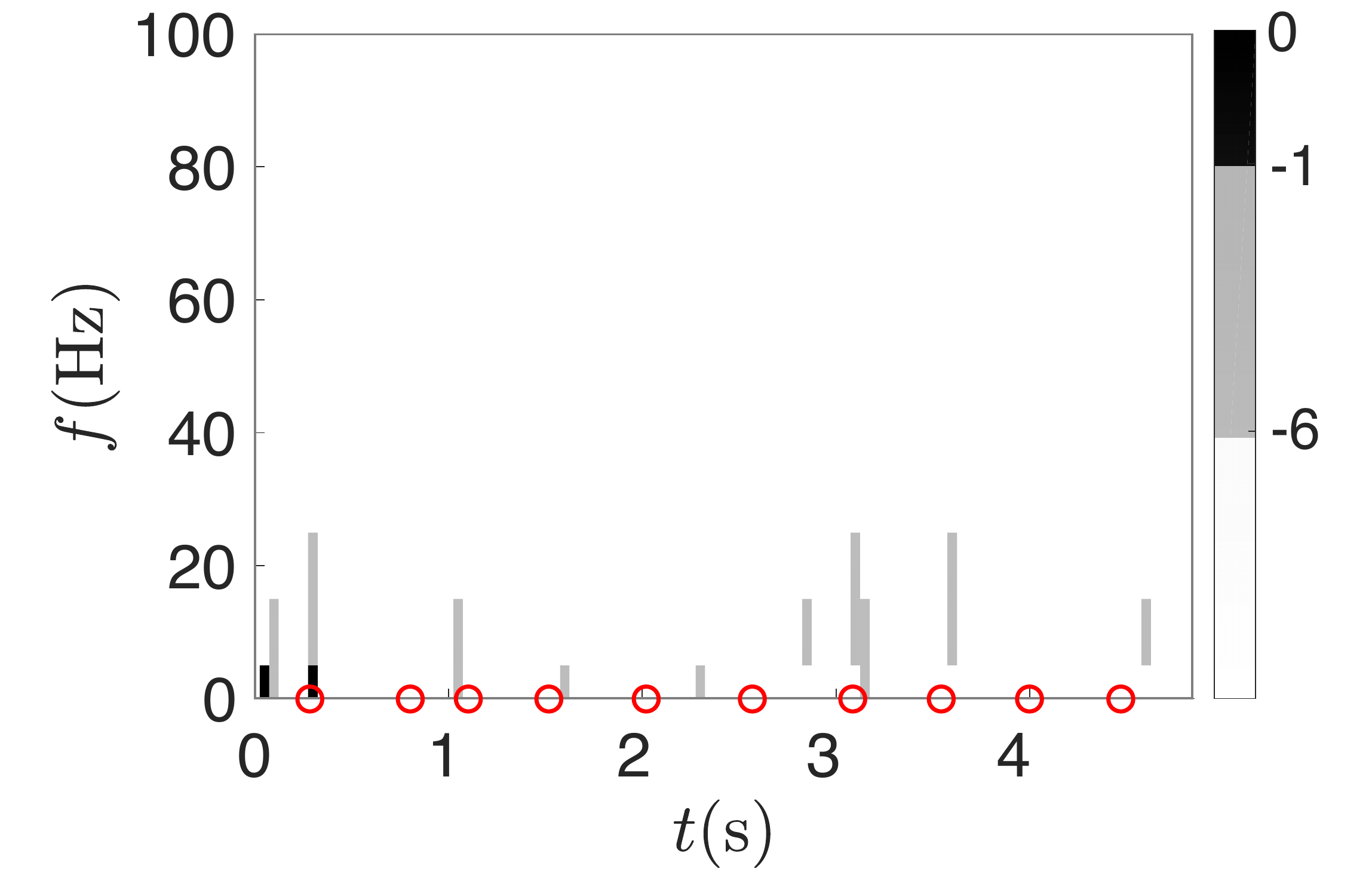}}
		\vspace{-1mm}
		\caption{Normalized signal energy [cf. (\ref{eq:nor_amplitude})] in dB for a temporal window   of 0.1 s. (a) and (c) report results during sinus rhythm. (b) and (d) report results during atrial fibrillation. In (c) and (d), black, grey, and white colors represent  energy levels  L1 (-1 ${\rm{ dB}} \le E \leq$ 0 ${\rm{ dB}}$), L2 (-6 ${\rm{ dB}} \le E <$ -1 ${\rm{ dB}}$), and L3 (E $\le$ -3 ${\rm{ dB}}$), respectively. Remark the harmonics of the 50 Hz disturbance in (a) and (b). The red circles mark the peak of the ventricular activity.}
		\label{fig:energy3}
		\vspace{-3mm}
	\end{figure}

	Figure \ref{fig:energy3} shows an example of normalized signal energy during sinus rhythm and atrial fibrillation. From Figures \ref{fig:energy3}(a) and \ref{fig:energy3}(b), we make two observations. First,  the ventricular activity energy is concentrated below 50 Hz, while this is not the case for the atrial activity. Second, higher-frequency components  have more energy during sinus rhythm than during atrial fibrillation. The latter is because the atrial activity during sinus rhythm changes more dramatically than during atrial fibrillation (see also Figure \ref{fig:data}).
	
	To better analyze these spectrograms, we discretized the energy in three levels labeled as L1 (-1 ${\rm{ dB}} \le E \leq$ 0 ${\rm{ dB}}$), L2 (-6 ${\rm{ dB}} \le E <$ -1 ${\rm{ dB}}$), and L3 ($E <$ -6 ${\rm{ dB}}$). 
	We illustrate an example  in Figures 4(c) and  4(d) during sinus rhythm and atrial fibrillation, respectively. The relevant frequency band for L1 is between 0 Hz and 20 Hz and for L2 is between 0 Hz and 80 Hz. These two temporal frequency bands are larger compared with the respective bands during atrial fibrillation. In other words, during sinus rhythm the signal has more energy in the higher temporal frequencies.

	In Table \ref{tb:fre}, we list the relevant temporal frequency bands of energy levels L1 and L2 for the ten patients. Although there is a slight variation of energy distribution among patients, overall we observed that  sinus rhythm signals have more energy in the higher frequencies compared with the atrial fibrillation signals. We found normalized energy larger than -1 dB  up to 50 Hz during sinus rhythm, while we found it only  up to 20 Hz during atrial fibrillation. Also, the frequency range of L2 during sinus rhythm  is wider than during atrial fibrillation. 
	
	\begin{table}
		\resizebox{0.48\textwidth}{!}{
			\begin{threeparttable}	
				\centering
				\caption{Relevant frequency bands of L1 and L2 normalized energy.}
				\begin{tabular}{  c c c c c}
					\toprule
					\multirow{2}{*}{Patient No.} & \multicolumn{2}{c}{Freq. bands of L1  (Hz)} & \multicolumn{2}{c}{Freq. bands of L2  (Hz)} \\ 
					\cmidrule(lr){2-3} \cmidrule(lr){4-5}
					&SR&AF&SR&AF\\
					\hline
					P1 & [0, 40] & [0, 10] & [0,  80] & [0, 30] \\
					
					P2 & [0, 40] & [0,  20] & [0, 110] & [0, 90] \\
					
					P3 & [0, 50] & [0,  20] & [0, 110] & [0, 50] \\
					P4 & [0, 50] & [0,  20] & [0, 100] & [0, 30] \\
					
					P5 & [0, 50] & [0,  20] & [0,  140] & [0, 30] \\
					
					P6 & [0, 40] & [0,  20] & [0,  90] & [0, 30] \\
					
					P7 & [0, 40] & [0,  20] & [0, 90] & [0, 40] \\
					
					P8 & [0, 20] & [0, 10] & [0, 80] & [0, 30] \\
					
					P9 & [0, 20] & [0, 10] & [0, 80] & [0, 50] \\
					
					P10 & [0, 50] & [0, 10] & [0, 70] & [0, 30] \\
					\hline
					mean & [0, 40] & [0, 16] & [0, 95] & [0, 41] \\
					
					std & [0, 11.55] & [0, 5.16] & [0, 20.68] & [0, 19.12] \\
					\bottomrule			
				\end{tabular}
				
				\label{tb:fre}
				\begin{tablenotes}
					\item L1: -1 ${\rm{ dB}} \le E \leq$ 0 ${\rm{ dB}}$;  L2: -6 ${\rm{ dB}} \le E <$ -1 ${\rm{ dB}}$;
					\item SR: sinus rhythm;  AF: atrial fibrillation;
				\end{tablenotes}
			\end{threeparttable}
		}
		\vspace{-3mm}
	\end{table}

	\subsection{GFT analysis}

	For the GFT analysis, we measured the normalized energy at different graph frequencies as
	\begin{equation} 
	E_\mathcal{G}(k,t)= 10 \text{log}_{10}(|\widetilde{y}_\text{n}(k,t)|^2)
	\label{eq:gftenergy}
	\end{equation} 
	where $|\widetilde{y}_\text{n}(k,t)|$ is the signal normalized amplitude [cf. (\ref{eq:nor_amplitude})]. The subscript $\mathcal{G}$ stresses that the analysis is in the graph frequency domain.

\begin{figure}[]%
		\centering
		\subfigure[]{%
			\includegraphics[width=4.2cm,height=2.5cm]{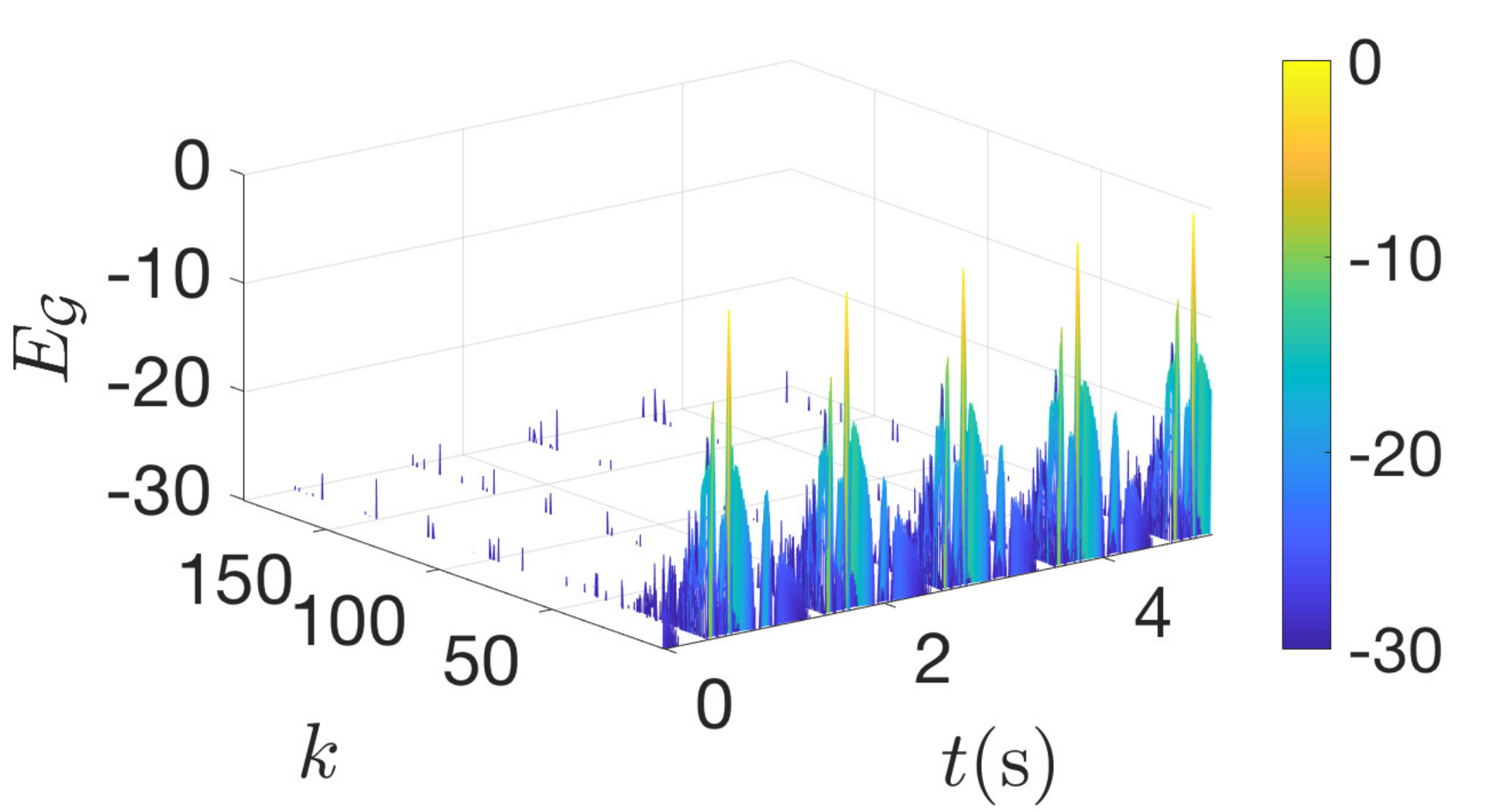}}%
		\hspace*{\fill}
		\subfigure[]{%
			\includegraphics[width=4.2cm,height=2.5cm]{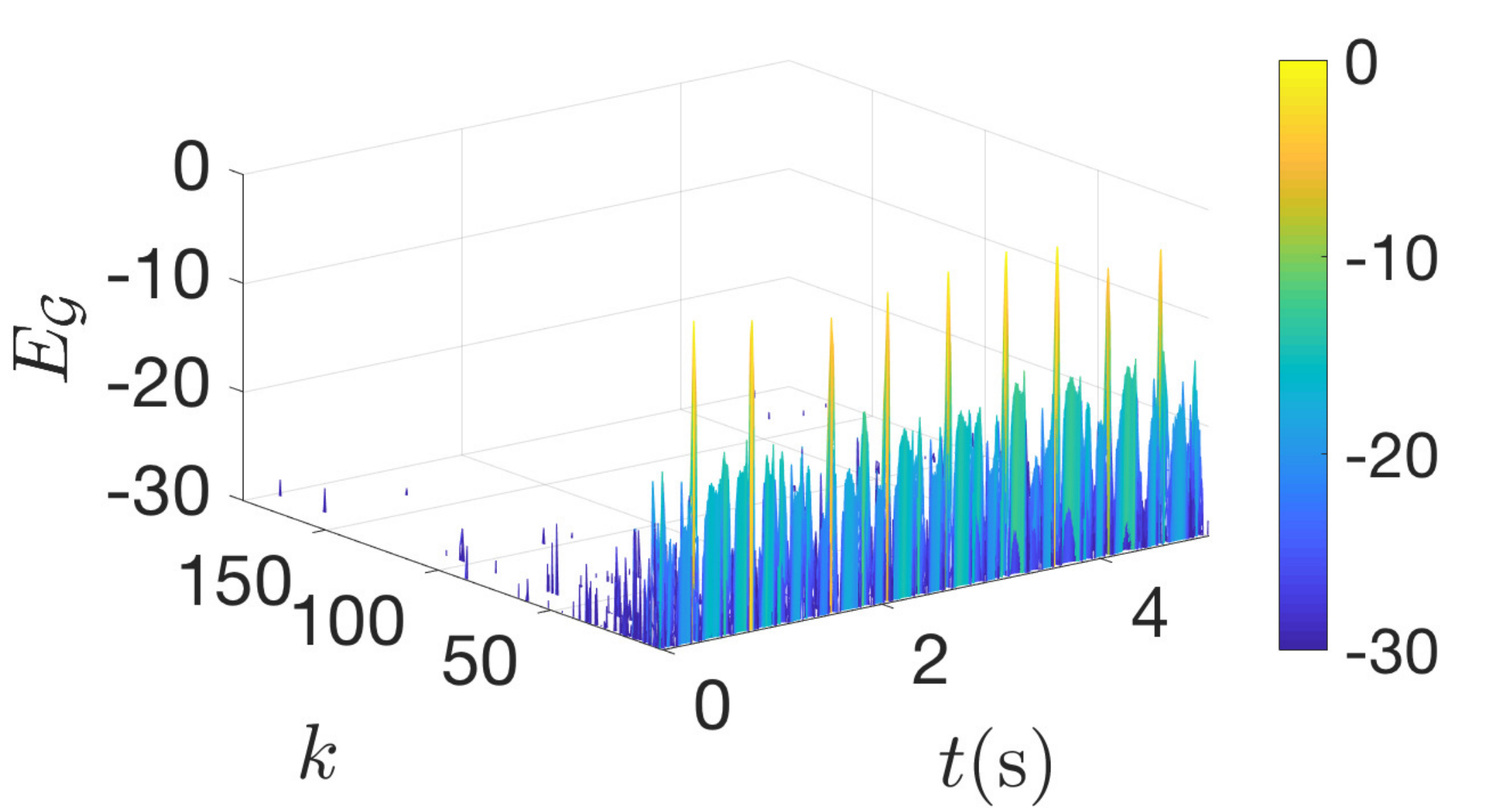}}%
		\caption{Normalized signal energy [cf. (\ref{eq:gftenergy})] in dB across time $t$ and graph frequency index $k$. (a)   sinus rhythm; (b)  atrial fibrillation.}
		\label{fig:GFT} 
	\end{figure}

	Figures \ref{fig:GFT}(a) and \ref{fig:GFT}(b) illustrate  respectively the normalized signal energy as a function of time and graph frequencies during sinus rhythm and atrial fibrillation for one patient. We found most of the signal energy concentrates in the low graph frequencies (i.e., lower $k$) and the signal energy decreases with the graph frequency (i.e., larger $k$).

\begin{figure}[]%
		\centering
		\subfigure[]{%
			\includegraphics[width=4.2cm,height=2.5cm]{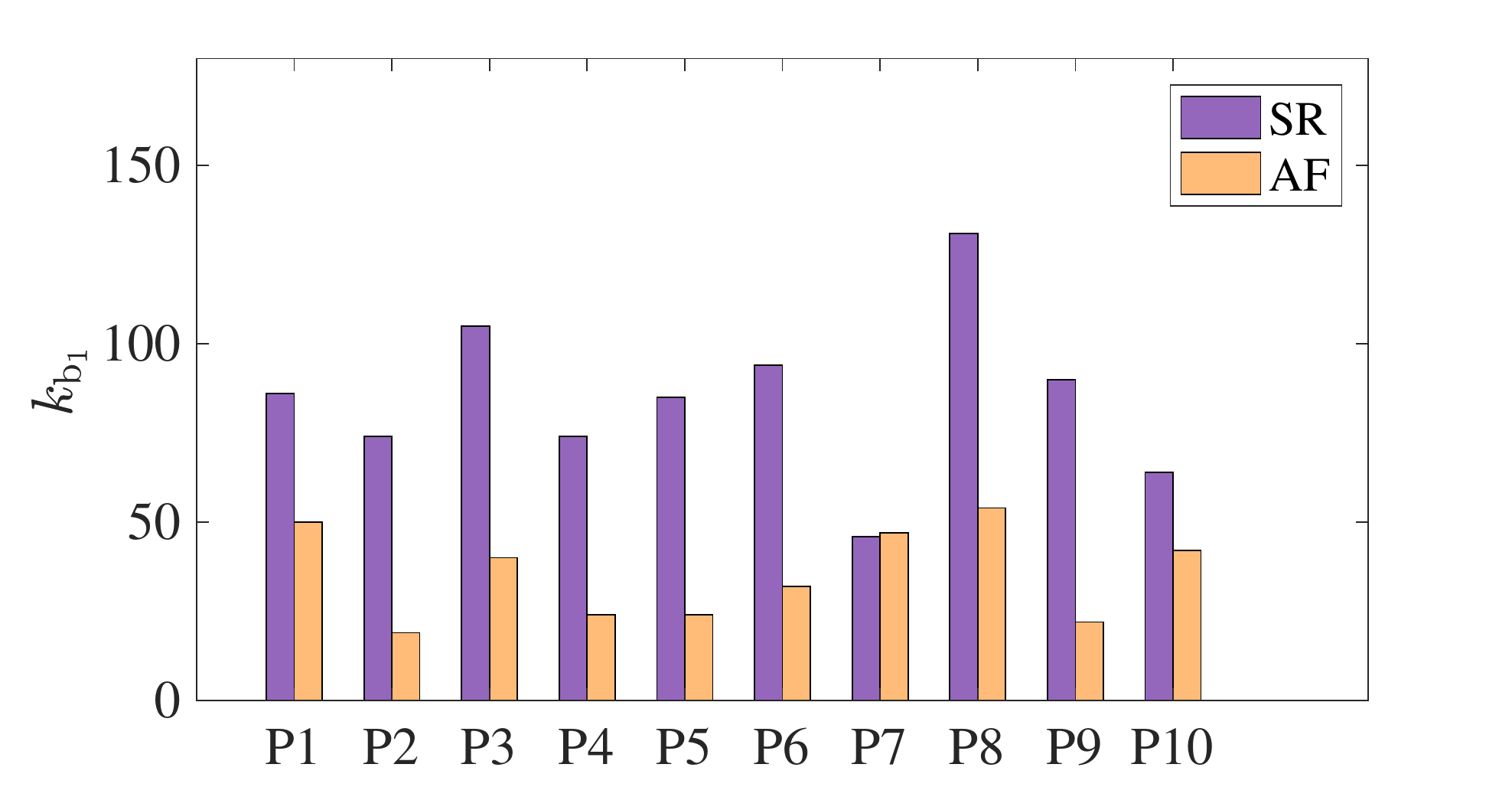}}%
		\hspace*{\fill}
		\subfigure[]{%
			\includegraphics[width=4.2cm,height=2.5cm]{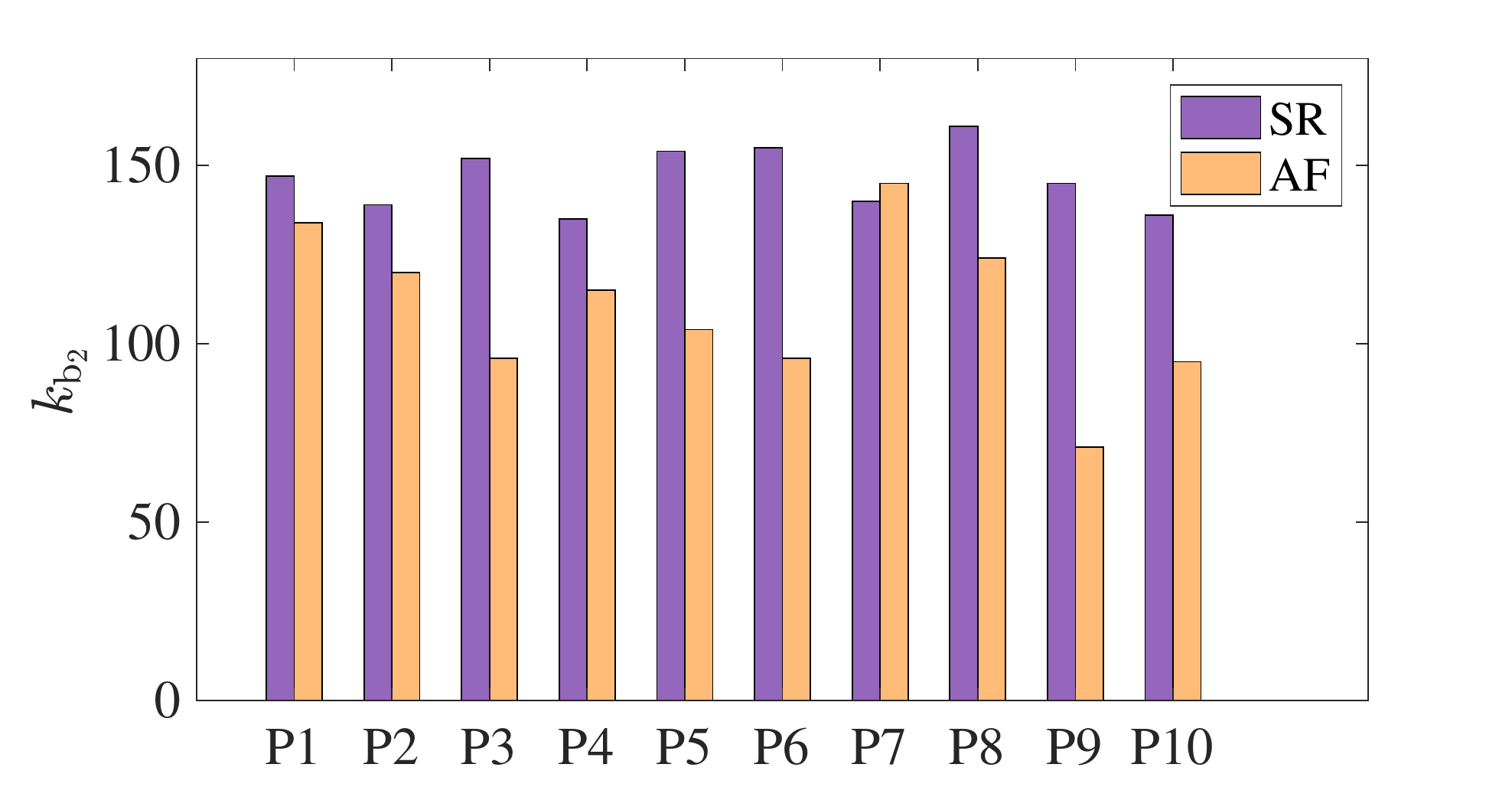}}%
		\caption{Boundary graph frequency indices for ten patients during sinus rhythm and atrial fibrillation. (a) graph frequency index $k_{b_1}$ (50\% energy concentrated in [0, $k_{b1}$]); (b) graph frequency index $k_{b_2}$ (80\% energy concentrated in [0, $k_{b1}$]).}
		\label{fig:boundary} 
	\end{figure}

	We further compare the energy distribution during sinus rhythm and atrial fibrillation for the ten patients. Since the energy decreases with the graph frequency (lower $k$), we considered two boundary graph frequencies $b_1$ and $b_2$ with indices $k_{b_1}$ and $k_{b_2}$, where 50\% and 80\% of the energy concentrates in the bands $[0, k_{b_1}]$ and $[0, k_{b_2}]$ for all time instants, respectively. Figures \ref{fig:boundary}(a) and \ref{fig:boundary}(b) compare the boundary graph frequency indices   during sinus rhythm and atrial fibrillation. The boundary graph frequencies during sinus rhythm are higher than during atrial fibrillation for almost all patients. This suggests that during sinus rhythm the EGM  has a larger graph bandwidth than during atrial fibrillation. That is, the signal changes faster across the graph (hence epicardium) during sinus rhythm than during atrial fibrillation.
	
	
\begin{figure}[]%
		\centering
		\subfigure[]{%
			\includegraphics[width=8.5cm,height=2.6cm]{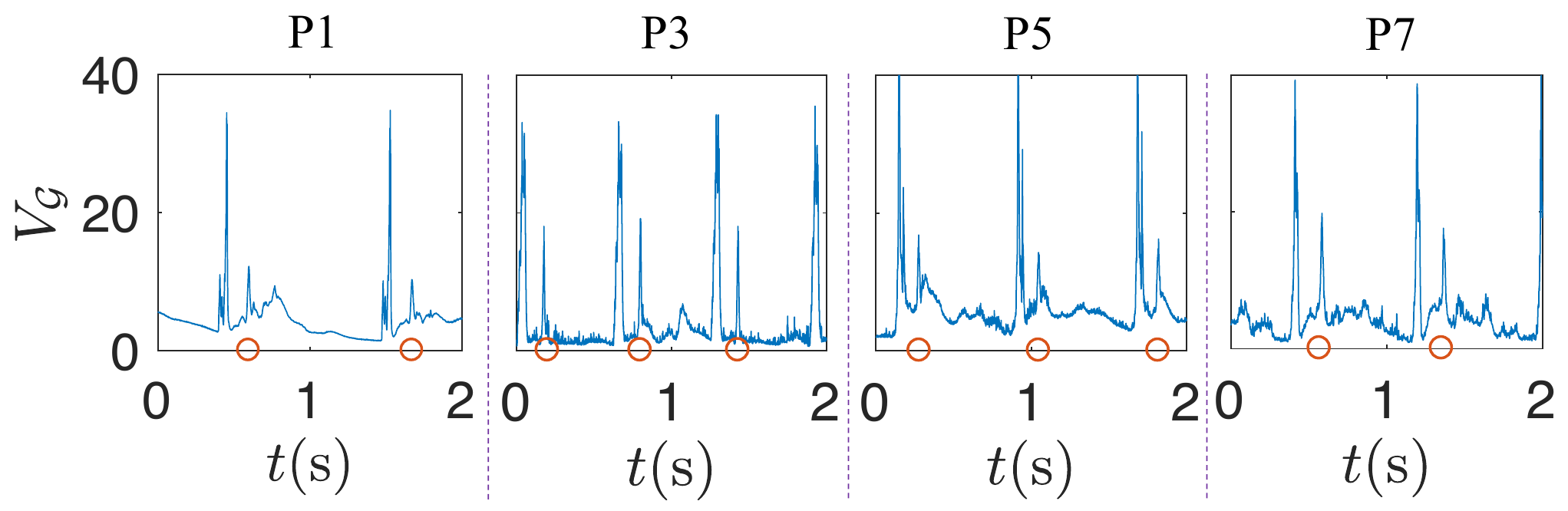}}%
		\qquad
		\subfigure[]{%
			\includegraphics[width=8.5cm,height=2.5cm]{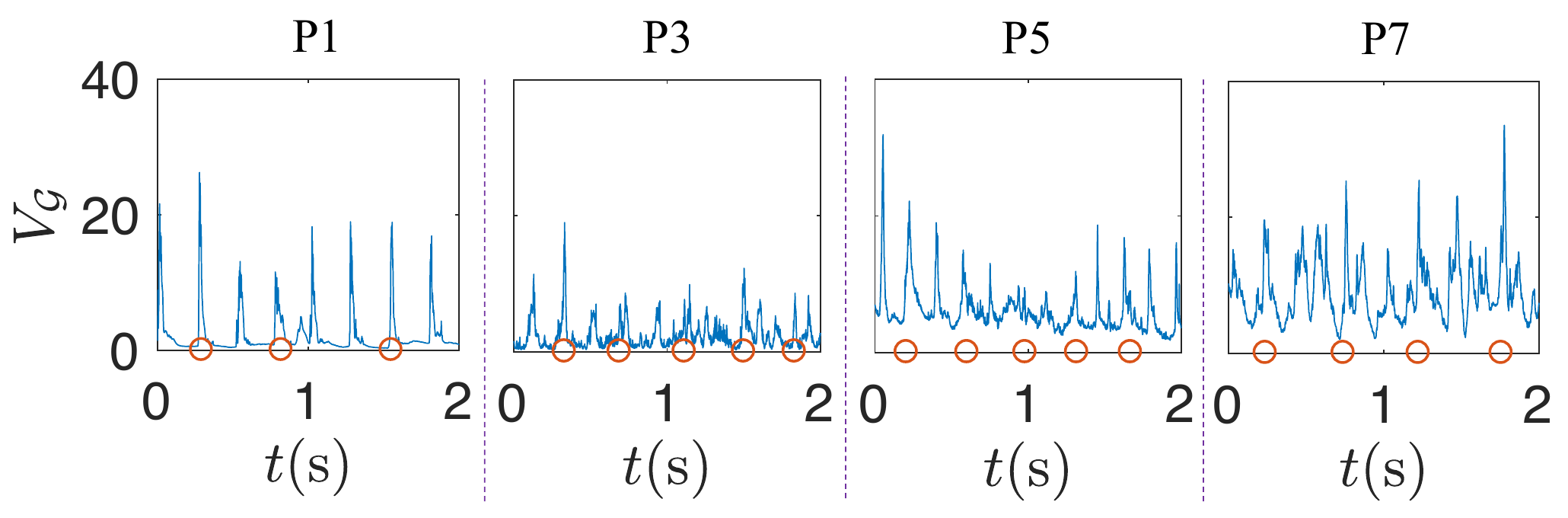}}%
		\caption{Graph smoothness measure of the  epicardial electrograms over the graph for four representative patients. (a) sinus rhythm; (b) atrial fibrillation. The red
			circles mark the peak of the ventricular activity.}
		\label{fig:variation}
	\end{figure}

	To analyze the signal smoothness  over the graph, we show in Figure \ref{fig:variation} the variation  [cf. (\ref{eq:tv})] for four representative patients. During sinus rhythm, the ventricular activity varies slower over the graph than the atrial activity. During atrial fibrillation, the  atrial activity overlaps with the ventricular activity, resulting in an increased variation during the ventricular rhythm; note the highest peaks in Figure \ref{fig:variation}(b). 
	
	We may expect a higher spatial variation of the atrial activity during atrial fibrillation than during sinus rhythm. This is because the signal changes more frequently across time during atrial fibrillation periods. However, as shown in Figure \ref{fig:variation}, the atrial activity has a larger spatial variation during sinus rhythm than during atrial fibrillation. To explain this counterintuitive  result, we need to exploit the association between the temporal and spatial variations.
	The spatial graph variation in (\ref{eq:tv}) measures only the EGM variation  per time instant and ignores the correlation across time. Since the temporal frequencies  provide additional insights on the EGMs and since the GFT alone does not capture them, we analyze next the EGMs  with the joint STFT and GFT to address the latter.

	\begin{figure}[]%
		\centering
		\subfigure[]{%
			\includegraphics[width=8.5cm,height=2.6cm]{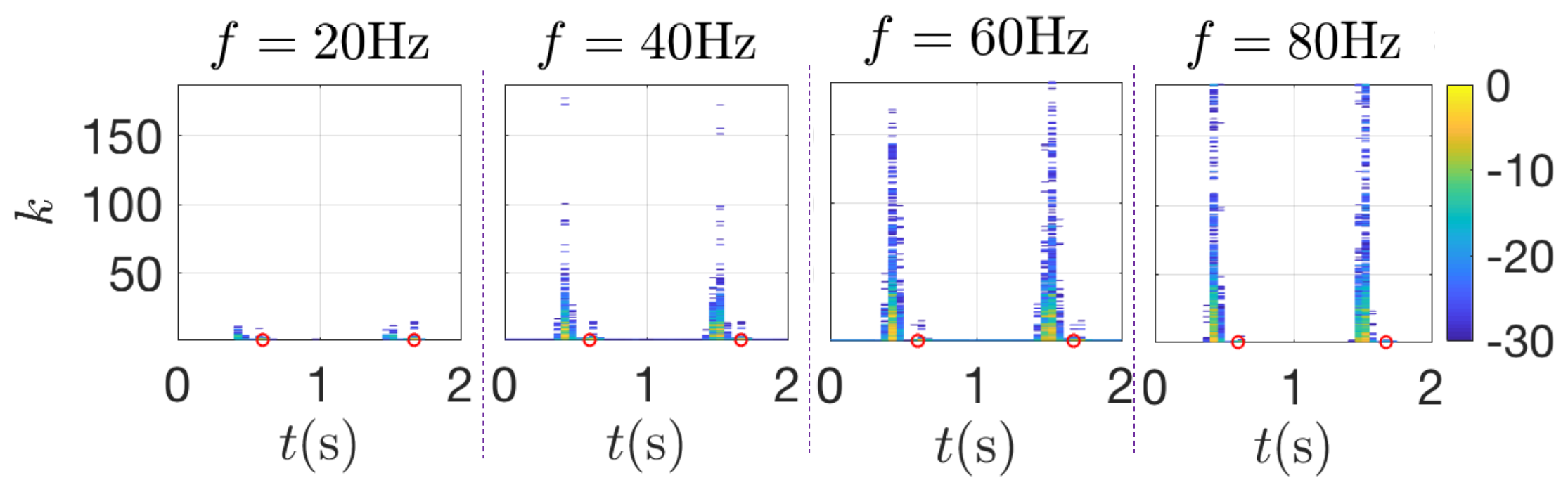}}%
		\qquad
		\subfigure[]{%
			\includegraphics[width=8.5cm,height=2.6cm]{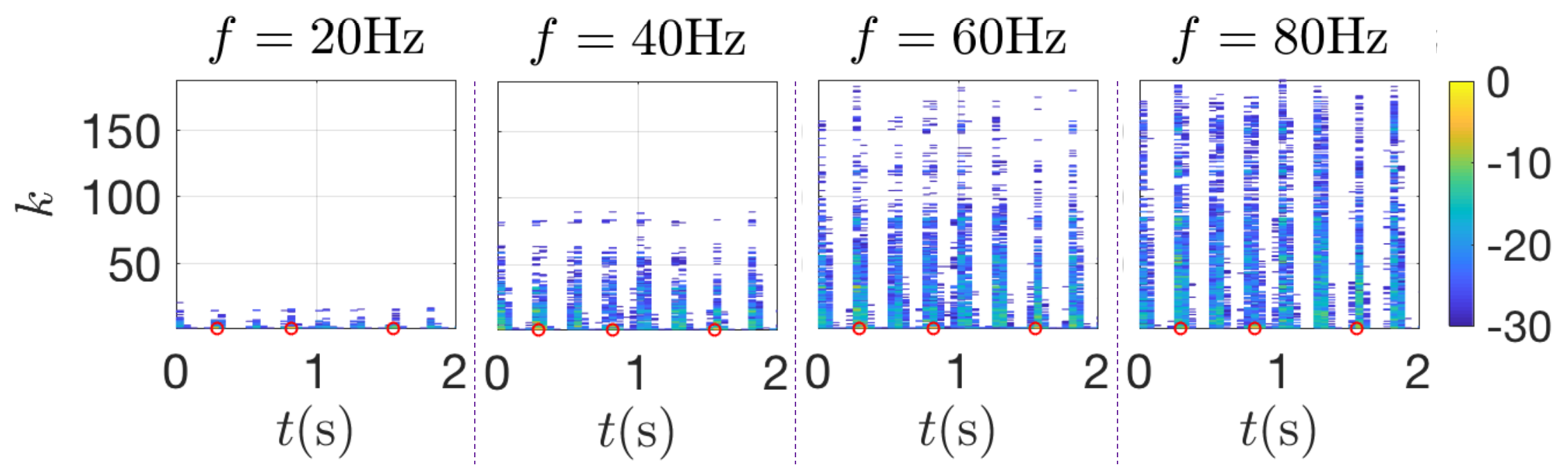}}%
		\caption{Normalized energy in dB in the joint graph and short-time Fourier transform domain. (a) sinus rhythm; (b) atrial fibrillation. The scalar $k$ represents the graph frequency index, $t(s)$ the time in seconds, and $f$ the temporal frequency. Each plot shows the spatial distribution of the signal energy as a function of time; different plots refer to different temporal frequencies.  The red
			circles mark the peak of the ventricular activity.}
		\label{fig:jgft}
	\end{figure}
	
	\subsection{Joint STFT and GFT analysis}
	We now analyze the normalized signal energy in the joint short-time Fourier transform and graph Fourier transform domain.
	
	Figures 8(a) and 8(b) depict the results during sinus rhythm and atrial fibrillation for one patient.
	To improve visualization, we focus on the temporal frequencies {20 Hz, 40 Hz, 60 Hz, and 80 Hz}. Overall, the temporal frequency components change slowly over the graph; this is reflected by the energy concentration in the low graph frequencies.
	However, we also observed that  higher temporal frequencies change faster over the graph compared with the lower ones; this is reflected by the higher energy concentration in the high graph frequencies for $f = 60$ Hz, and $80$ Hz.

	\begin{figure}[]%
		\centering
		\subfigure[]{%
			\includegraphics[width=8.5cm,height=2.6cm]{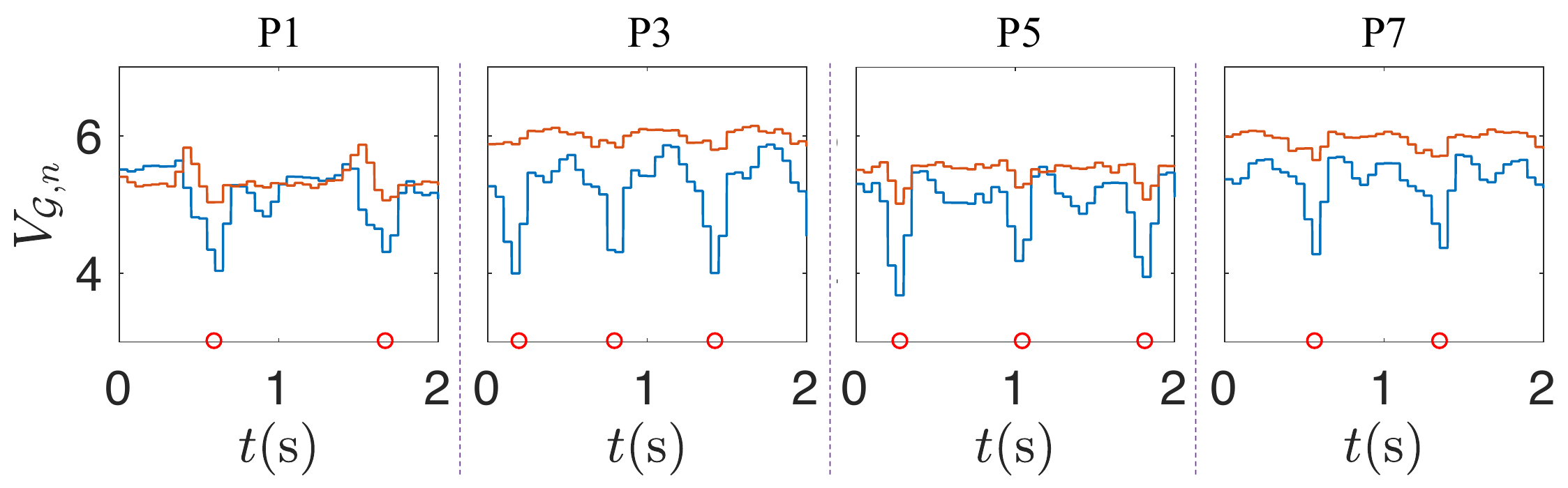}}%
		\qquad
		\subfigure[]{%
			\includegraphics[width=8.5cm,height=2.6cm]{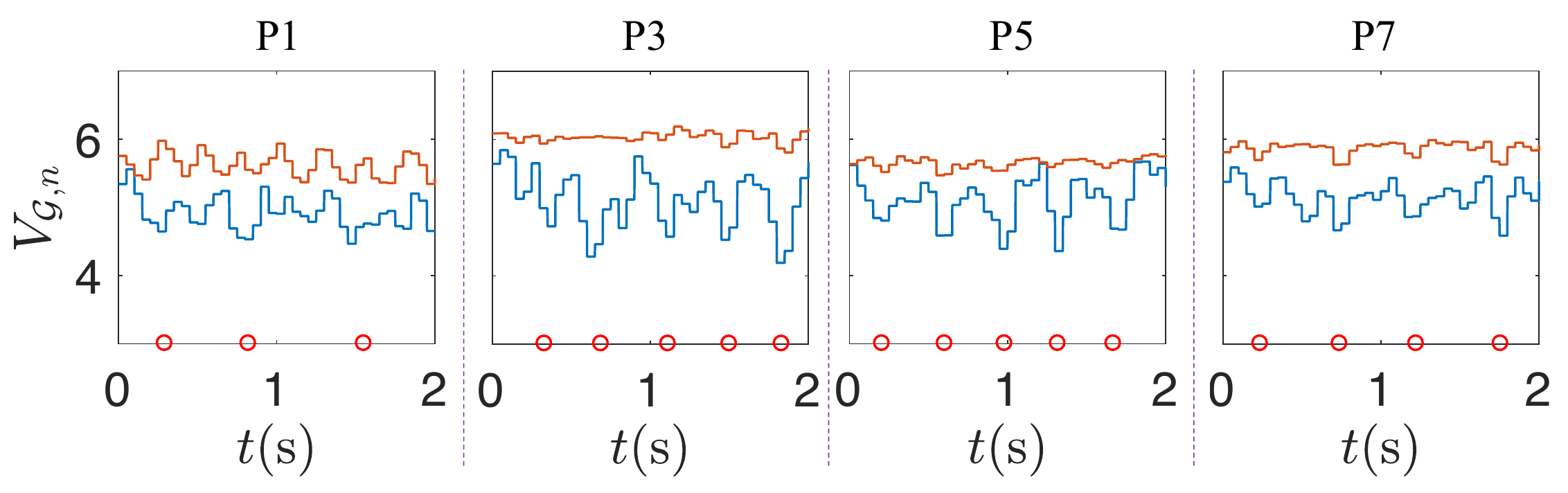}}%
		\caption{Smoothness measure over time of the low and high temporal frequencies in the joint graph and short-time Fourier transform domain. (a) sinus rhythm; (b) atrial fibrillation. The red and blue lines indicate the mean smoothness of the low and high temporal frequencies, respectively. The red
			circles mark the peak of the ventricular activity.}
		\label{fig:JGFT_V}
		\vspace{-1mm}
	\end{figure}

	To quantify the graph spatial variations of the low (0 Hz to 100 Hz) and high (100 Hz to 500 Hz) temporal frequencies, we calculated the average variation following (\ref{eq:tvn}). Due to space limitation, we  show in Figure \ref{fig:JGFT_V} the results for  four representative patients. The high temporal frequencies have a larger graph variation compared to the lower temporal frequencies. From the STFT analysis (Table \ref{tb:fre}), the EGM has more energy in the high temporal frequencies during sinus rhythm than  during atrial fibrillation. This explains the result in the GFT analysis (Figure \ref{fig:variation}), i.e., the atrial activity during sinus rhythm has a higher spatial variation than  during atrial fibrillation. This suggests that   the spatial variation is correlated to the temporal variation. If a signal changes rapidly across time, it will have higher energy in the high temporal frequencies. This high variation across time translates then into a higher variation over the graph.
	
	During sinus rhythm, the average spatial variation  decreases to a small value when the ventricular activity appears. That is, the temporal frequencies change slower over the atria during  the ventricular activity than during the  atrial activity. But during atrial fibrillation, the   spatial variation during the ventricular rhythm is higher because of the coupling between the atrial  and  ventricular activities. 
	
	The above analysis shows that it is  possible to separate the atrial and ventricular activities based on their spatial variations. This separation would be infeasible by  the STFT alone (which ignores correlation across space) or by the GFT alone (which ignores correlation across time). Since the joint transform analyzes the graph signal in  short-time periods, it improves separation of the two activities in the joint domain. In the next section, we will leverage these observations to extract the  atrial activity  in the joint domain. 
	
	\section{Atrial Activity Extraction}
	Recall that the atrial activity   measurements are often corrupted by ventricular activity. In the sequel, we propose an algorithm to extract atrial activity from the mixed measurements   based on the graph and time variations of the atrial and ventricular activities.
	
	\subsection{Algorithm}

	The graph-time analysis in Section IV-C showed that   the ventricular activity is smoother over the graph than the atrial activity. We, therefore, exploit the difference in smoothness to estimate the ventricular activity from the noisy epicardial measurement. The atrial activity can be then obtained by subtracting the estimated ventricular activity from the EGM.
	
	By considering the EGM as a linear combination of the atrial activity and the ventricular activity \cite{rieta2007comparative}, we can write  the mixed signal $\mathbf{y}(t)$ over the $K$ electrodes at time $t$ as
	\begin{equation}
	\mathbf{y}(t) = \mathbf{a}(t)  + \mathbf{v}(t)
	\end{equation}
	where $\mathbf{a}(t)$ indicates the atrial signal and $\mathbf{v}(t)$  the ventricular  signals.
	By applying enframing (segmenting the signal into overlapping frames), we represent the signal at frame  $\tau$ in the matrix form as
	\begin{equation}
	\mathbf{Y}(\tau) = \mathbf{A}(\tau) +\mathbf{V}(\tau)
	\label{eq:eq16}
	\end{equation}
	where  $\mathbf{Y}(\tau)$, $\mathbf{A}(\tau)$, and $\mathbf{V}(\tau)$ are $K \times T_{M}$ matrices following from  (\ref{Eq:time_sample_matrix}).
	Then, from the joint STFT and GFT transform we get the joint spectral representation
	\begin{equation}
	\widetilde{\mathbf{Y}}(\tau) = \widetilde{\mathbf{A}}(\tau)  +\widetilde{\mathbf{V}}(\tau)
	\end{equation}
	where $\widetilde{\mathbf{Y}}(\tau)$, $\widetilde{\mathbf{A}}(\tau)$, and  $\widetilde{\mathbf{V}}(\tau)$
	are the joint transforms of the mixed EGM signal, atrial activity, and ventricular activity, respectively.  The respective columns are $\widetilde{\mathbf{y}}(\tau,f)$, $\widetilde{\mathbf{a}}(\tau,f)$, and $\widetilde{\mathbf{v}}(\tau,f)$.
	
	Since the ventricular activity $\widetilde{\mathbf{v}} (\tau,f)$ is smoother than the atrial activity $\widetilde{\mathbf{a}} (\tau,f)$, we estimate $\widetilde{\mathbf{v}} (\tau,f)$ as a smooth graph signal reconstruction with 
	minimum distortion from the mixed signal $\widetilde{\mathbf{y}}(\tau,f)$. Mathematically, this consists of solving the problem
	\begin{mini}|l|
		{\widetilde{\mathbf{v}}(\tau,f)}{||\widetilde{\mathbf{y}}(\tau,f)-\widetilde{\mathbf{v}}(\tau,f)||_{2}^{2} }{}{}
		\addConstraint{\frac{\widetilde{\mathbf{v}}^{H}(\tau,f)\mathbf{\Lambda}\widetilde{\mathbf{v}}(\tau,f)}{\widetilde{\mathbf{v}}^{H}(\tau,f)\widetilde{\mathbf{v}}(\tau,f)}\leqslant  c}
		\label{eq:opt}.
	\end{mini}
	where the cost function seeks for finding a ventricular signal   $\widetilde{\mathbf{v}} (\tau,f)$ that is close to the EGM measurement $\widetilde{\mathbf{y}} (\tau,f)$, while the constraint imposes the maximum normalized variation to be at most $c$ for all frames $\tau$ and temporal frequencies $f$.
	
	By rearranging (\ref{eq:opt}) as
	\begin{mini}|l|
		{\widetilde{\mathbf{v}}(\tau,f)}{||\widetilde{\mathbf{y}}(\tau,f)-\widetilde{\mathbf{v}}(\tau,f)||_{2}^{2} }{}{}
		\addConstraint{\widetilde{\mathbf{v}}^{H}(\tau,f)(\mathbf{\Lambda}-c\mathbf{I})\widetilde{\mathbf{v}}(\tau,f)}\leqslant 0
	\end{mini}
	and  defining the Lagrangian
	\begin{subequations}
		\begin{align}
		L(\widetilde{\mathbf{v}}(\tau,f) , \mu, c) &=  ||\widetilde{\mathbf{y}}(\tau,f)-\widetilde{\mathbf{v}}(\tau,f)||_{2}^{2} + \mu g(\mathbf{\widetilde{\mathbf{v}}},c)\\
		g(\mathbf{\widetilde{\mathbf{v}}},c)&=\widetilde{\mathbf{v}}^{H}(\tau,f)(\mathbf{\Lambda}-c\mathbf{I})\widetilde{\mathbf{v}}(\tau,f)
		\label{eq:opt1}
		\end{align}
	\end{subequations}
	we can find the ventricular activity $\widetilde{\mathbf{v}}(\tau,f)$ by solving the Karush-Kuhn-Tucker conditions
	
	\begin{equation}
	\begin{aligned}
	\frac{\partial  L(\widetilde{\mathbf{v}}(\tau,f), \mu, c)}{\partial \widetilde{\mathbf{v}}(\tau,f)} &=\mathbf{0}\\
	\mu  g(\mathbf{\widetilde{\mathbf{v}}}(\tau,f),c) &= 0 \\
	\mu & \geqslant 0,
	\end{aligned}
	\end{equation}
	where $\mu$ is the Lagrangian multiplier \cite{boyd2004convex}. The closed-form solution to (\ref{eq:opt}),  i.e., the estimated ventricular activity,  is given by
	\begin{equation}
	\widetilde{\mathbf{v}}(\tau,f) = [(1-\mu c)\mathbf{I}+\mu \mathbf{\Lambda}]^{-1}\widetilde{\mathbf{y}}(\tau,f).
	\label{eq:extraction}
	\end{equation}
	After estimating the ventricular activity, we can recover the atrial activity by
	\begin{equation}
	\widetilde{\mathbf{a}}_{\text{est}}\left( \tau, f \right)=\widetilde{\mathbf{y}}\left(  \tau, f \right) -  \widetilde{\mathbf{v}}\left( \tau, f \right).
	\end{equation}
	Finally, we obtain the time domain signals through the inverse transforms.

	The proposed   algorithm relies on the presence of the ventricular activity. Since the ventricular activity has most of its energy in the zero graph frequency (see Figure 8), we can detect it by thresholding the energy in the joint STFT and GFT domain. If the energy in the zero graph frequency index ($k = 0$) exceeds this threshold, it indicates the presence of the ventricular activity.

	\begin{figure*}
		\centering
		\includegraphics[width=16cm,height = 4.8cm]{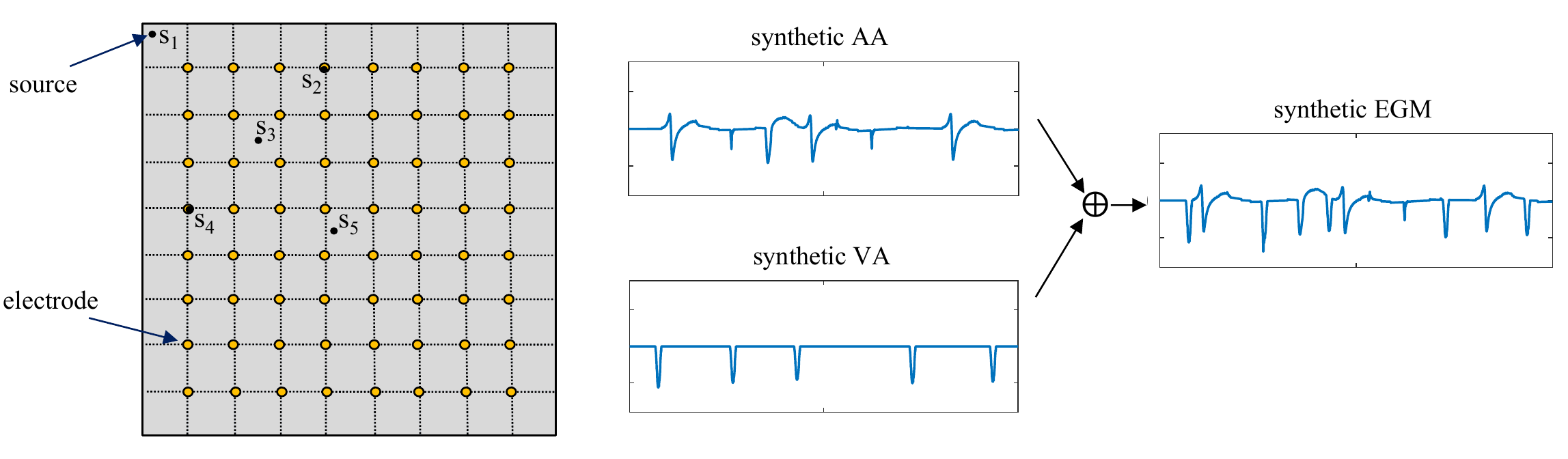}
		
		\caption{Simulation set up and synthetic signals during atrial fibrillation. \textit{Left}: simulated two-dimensional tissue with $8 \times 8$ electrodes on top of it. Five foci sources $s_1$ to $s_5$ initiate the atrial fibrillation. \textit{Right}: An example of synthetic atrial activity (AA), ventricular activity (VA), and mixed epicardial electrogram (EGM) with an atrial cycle length of 160 ms.}
		\label{fig:Simulation}
	\end{figure*}

	\subsection{Evaluation}
	
	To evaluate the performance of the proposed graph-based atrial activity extraction (GAE) algorithm, we need the ground truth pure atrial activity. However, this is unknown for real measurements; hence, we first evaluate the GAE algorithm with sythetic signals. We defer the test with real EGMs for the second part of this section. We compared the GAE algorithm with three popular alternatives:   average beat subtraction (ABS) \cite{slocum1985computer};   adaptive ventricular cancellation (AVC) \cite{rieta2007comparative}; and   independent component analysis (ICA) \cite{rieta2007comparative}.
	
	{\textit{Synthetic data generation:}}
	There exists several methods to simulate the atrial activity, see e.g., \cite{rieta2000atrial, stridh2001spatiotemporal, castells2005spatiotemporal, castells2003atrial}. 
	These algorithms   simulate well the electrogram during sinus rhythm, but face difficulties to simulate the atrial fibrillation electrogram. This is because of the overlap between the atrial and ventricular activities. Also, these methods are more suitable to generate  body surface ECGs rather than EGMs. The work in \cite{corino2013ventricular} generates  atrial EGMs by simulating the activation of the atrial fibers from the movement of a single dipole, which is less  realistic. In this work we focus on the atrial cell level to model the action potential during atrial fibrillation and extend it to the two-dimensional monodomain tissue. The atrial fibrillation is driven by the so-called ectopic foci sources that are located in various points of the tissue. This is one of the standard atrial fibrillation mechanisms in advanced research \cite{haissaguerre1998spontaneous, ganesan2017simulation}.
	
	The cell action potential follows the Courtemanche model of human atrial cells \cite{courtemanche1998ionic}. 
	To simulate the atrial activity during atrial fibrillation, we reduced the ionic conductance of $I_{\text{to}}$ to 50\%, $I_{\text{Kur}}$ to 50\% and $I_{\text{CaL}}$ to 30\% \cite{courtemanche1999ionic}. This is  based on the experimental study of chronic atrial fibrillation in \cite{courtemanche1999ionic}.
	After generating the signal at the cell level, we used the reaction-diffusion equation to simulate the propagation of the action potential  along the tissue \cite{plonsey2007bioelectricity}. The diffusion equation is given by
	\begin{equation}
	C_{\text{m}}\frac{\partial V_{\text{m}}}{\partial t} =I_{\text{tm}}+I_{\text{stim}}-I_{\text{ion}},
	\label{eq:diff}
	\end{equation}
	where $V_{\text{m}}$ is the transmembrane potential, $C_{\text{m}}=100$ pF is the transmembrane capacitance,  $I_{\text{ion}}$ is the total ionic current calculated from the Courtemanche model, $I_{\text{stim}}$ is the stimulus current, and $I_{\text{tm}}$ is the transmembrane current. The latter is calculated as
	\begin{equation}
	I_\text{tm} = \frac{1}{S_v}\nabla \cdot(\mathbf{D} \nabla V_{\text{m}}),
	\end{equation}
	where $S_v$ is the surface-to-volume ratio, $\nabla(\cdot)$ is the partial derivative operator, and $\mathbf{D}$ is the conductivity tensor.
	
	We considered a two-dimensional tissue   of 200 $\times$ 200 cells with a cell radius of 5 $\mu$m. The  longitudinal conductivity  is   100 cm/s. The transversal to longitudinal conductivity ratio is one-to-two. We discretized the model  through finite differences with resolution 0.01 cm and solved the reaction-diffusion equation [cf. (\ref{eq:diff})] with the Euler method with a  time step of 0.05 ms. Five ectopic foci sources drove the irregular atrial activity as illustrated in Figure 10.   We apply stimuli of 50 ms in length on these positions. Two atrial cycle length of 160 ms and 180 ms are used to simulate different degrees of atrial fibrillation. For each type, we generated six segments of 10s each.
	
	After generating the atrial activity, the next step is to generate the ventricular activity. The ventricular morphology is obtained  by cutting out the ventricular segment in a heart beat during real sinus rhythm \cite{castells2003atrial}. We inserted local variations  in the amplitude and width of the different ventricular segments. Finally, we added the ventricular activity to the synthetic atrial activity  to generate the mixed EGM.

	Given the high computational complexity of these simulations, we considered an array of only 8$\times$8 electrodes with the same inter-electrode spacing as the mapping array in Figure 1. The array is put on the  tissue to measure the atrial EGM. The  atrial EGM $\Phi(\mathbf{z},t)$ measured by the electrode at location $\mathbf{z}$ at time $t$ is calculated by \cite{virag2002study}
	\begin{equation}
	\Phi(\mathbf{z},t) = \frac{1}{4\pi \sigma_e}\int \frac{I_\text{tm}}{||\mathbf{z}-\mathbf{x}||}\text{d}\mathbf{x}
	\end{equation}
	where $\mathbf{z}$ and $\mathbf{x}$  represent the location vectors of the electrode and the cell, respectively,  and $\sigma_e$ is the extra-cellular conductivity.

	\begin{figure*}
		\centering
		\includegraphics[width=17cm,height = 4cm]{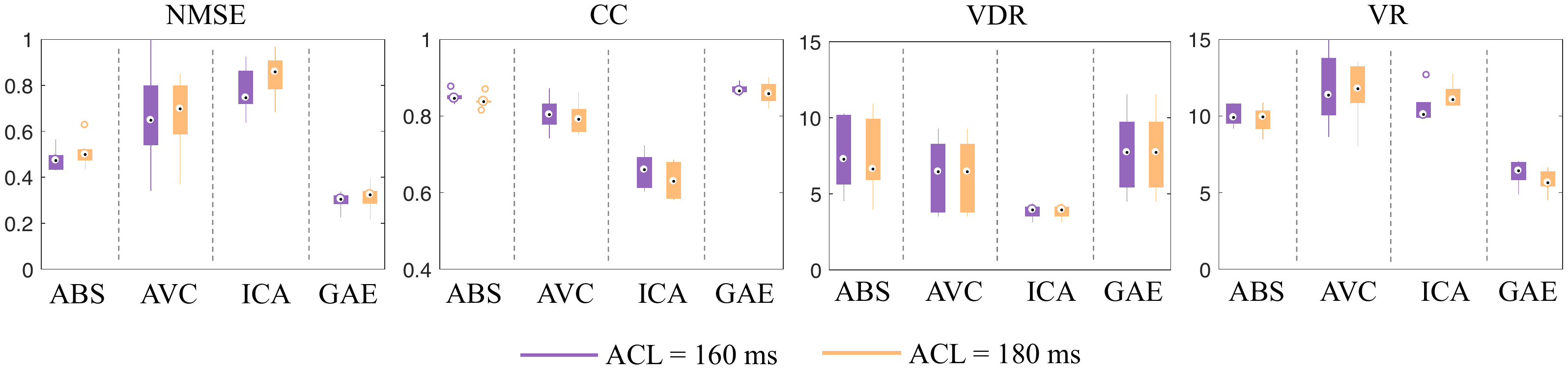}
		
		\caption{Boxplot  performance on  synthetic data of the  average beat subtraction (ABS) \cite{slocum1985computer},  adaptive ventricular cancellation (AVC) \cite{rieta2007comparative}, independent component analysis (ICA) \cite{rieta2007comparative}, and the proposed graph-based atrial activity extraction (GAE) method. Two atrial cycle length (ACL) of 160
			ms and 180 ms are considered.  The proposed GAE method achieves the lowest NMSE and VR, and highest CC and VDR in both cases. The   boxplots of NMSE, CC, and VR for the GAE method are comparatively short, which suggest that the GAE performance is more stable. Similar condensed boxplots are also observed for the ABS, but it presents outliers in the plots of NMSE and CC. }
		\vspace{-0.4cm}
		\label{fig:synthesize}
	\end{figure*}

	\begin{figure}[]%
		\centering
		\includegraphics[width=8.5cm,height=3.3cm]{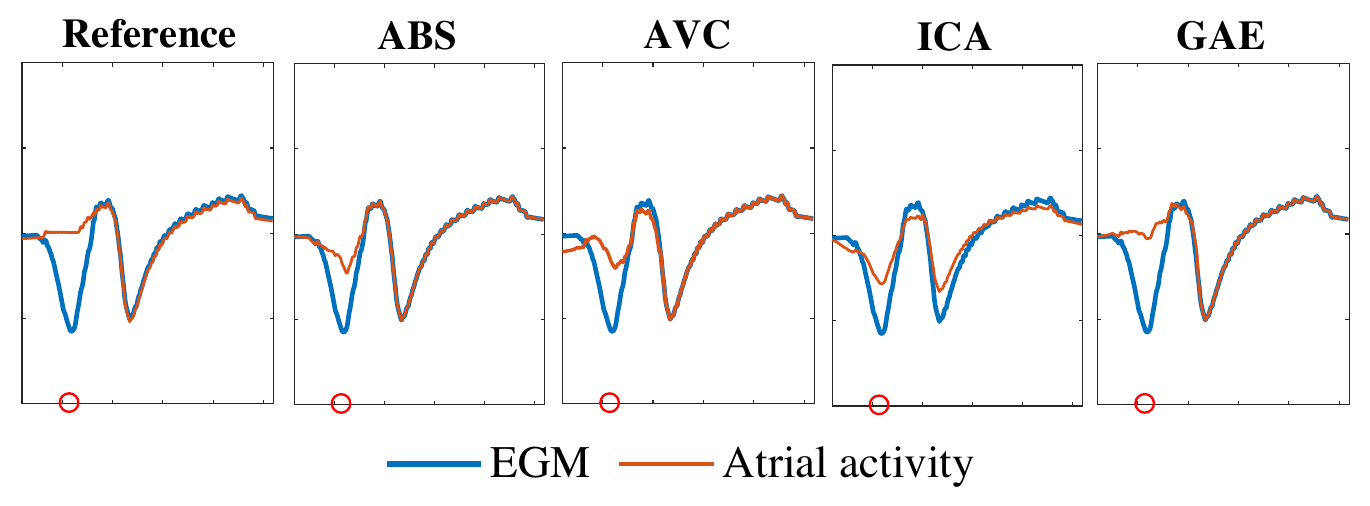}
		\caption{Illustrative example of the synthetic epicardial electrogram (EGM),   synthetic pure atrial activity (AA) and the estimated atrial activity by the different algorithms. The left plot shows the synthetic EGM (blue) and the synthetic atrial activity (red). The other plots show the synthetic EGM (blue) and the estimated atrial activity (red) with different algorithms: average beat subtraction (ABS) \cite{slocum1985computer};  adaptive ventricular cancellation (AVC) \cite{rieta2007comparative}; independent component analysis (ICA) \cite{rieta2007comparative}; proposed graph-based atrial activity extraction (GAE). The red circles mark the peak of the ventricular activity determined by the ECG measurements.}
		\vspace{-0.3cm}
		\label{fig:ext1}   
	\end{figure}
	
	\begin{figure}[]%
		\centering
		\includegraphics[width=8.5cm,height=7cm]{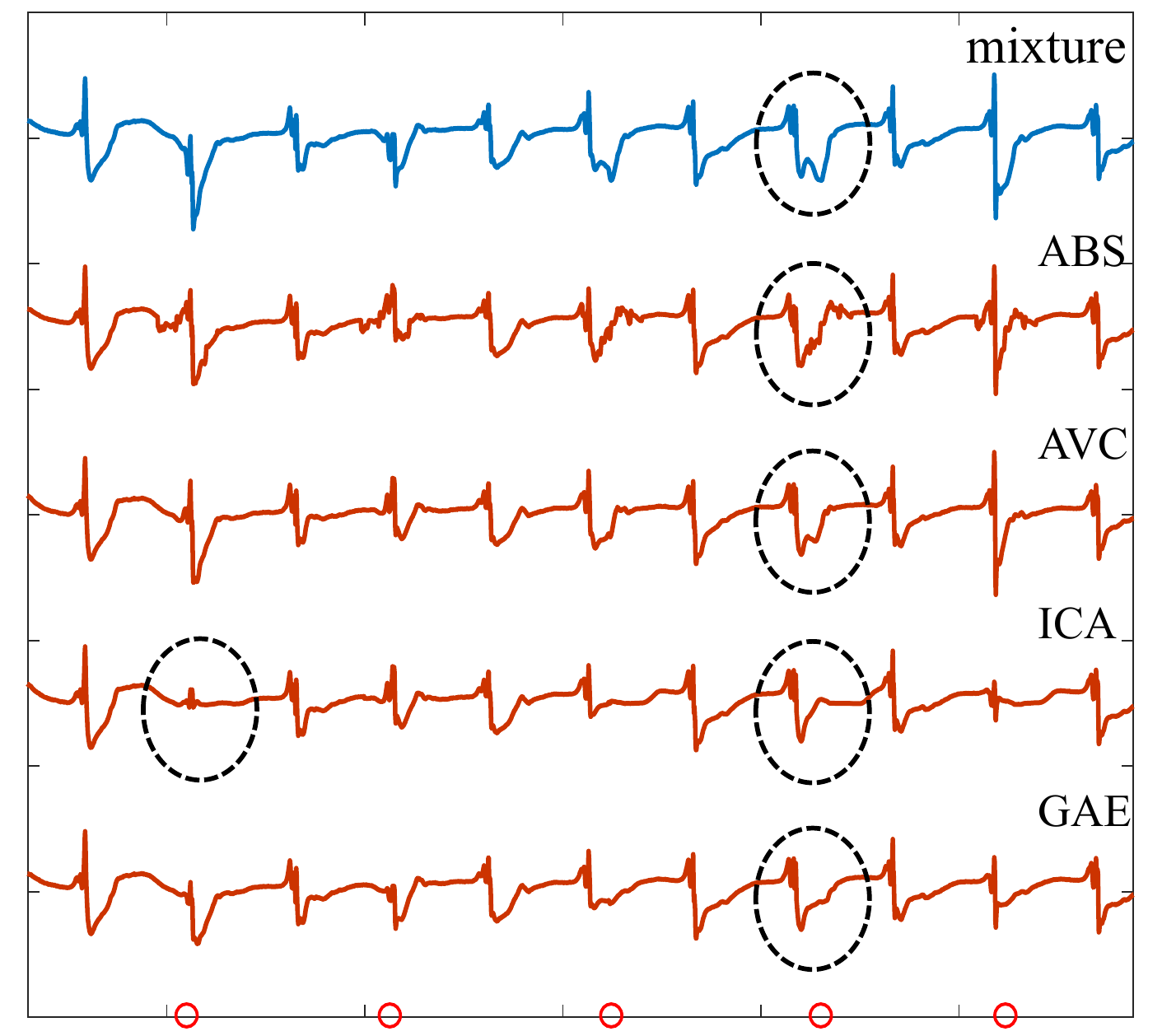}
		\caption{Illustrative example of the clinical epicardial electrogram (EGM) (blue) and the extracted atrial activity (red) by different algorithms:   average beat subtraction (ABS) \cite{slocum1985computer};  adaptive ventricular cancellation (AVC) \cite{rieta2007comparative}; independent component analysis (ICA) \cite{rieta2007comparative}; proposed graph-based atrial activity extraction (GAE). The proposed GAE method has  less fluctuations and distortions and removes more ventricular activity.}
		\vspace{-0.3cm}
		\label{fig:ext2}   
	\end{figure}

	{\textit{Performance metrics:}}
	In the synthetic scenario, we compared the estimated atrial activity with the pure atrial activity in terms of the normalized mean square error (NMSE) and the cross-correlation coefficient (CC). The NMSE is defined as
	\begin{equation}
	\text{NMSE} = \frac{1}{K}\sum_{i=1}^{K} \left(\frac{\sum_{t=0}^{T-1}(a_i(t)-a'_i(t))^2}{\sum_{t=0}^{T-1}(a_i(t))^2}\right)
	\end{equation}
	where $T$ is the length of the estimated atrial signal in the time domain, $a_i(t)$ and $a'_i(t)$ are the pure and the estimated atrial signals of the $i$th electrode at time $t$, respectively. The NMSE measures the normalized difference between the pure and the estimated atrial signals averaged over $K$ electrodes: a lower value  indicates a better estimation.
	
	The cross-correlation coefficient  is defined as
	
	\begin{small}
		\begin{equation}
		\text{CC} = \frac{1}{K}\sum_{i=1}^{K}\left(\frac{\sum_{t=0}^{T-1}\left(a_i(t)-\bar{a}_i\right)\left(a'_i(t)-\bar{a}'_i\right)}{\sqrt{\sum_{t=0}^{T-1}\left(a_i(t)-\bar{a}_i\right)^2}\sqrt{\sum_{t=0}^{T-1}\left(a'_i(t)-\bar{a}'_i\right)^2}}\right)
		\end{equation}
	\end{small}
	
	{\noindent}where $\bar{a}_i$ and ${\bar{a}'_i}$ are the mean of the pure and the mean of the estimated atrial signals of the $i$th electrode, respectively. The CC measures the similarity between the pure and the estimated atrial signals  averaged over $K$ electrodes: it is close to one if the pure and estimated atrial activities are correlated, and it is close to zero  otherwise.
	
	In the real EGM scenario, it is impossible to use intrusive measures to quantify algorithm performance through NMSE and CC since the ground truth  is unknown. Hence, we use two non-intrusive metrics, namely: the  ventricular depolarization reduction (VDR) \cite{rieta2007comparative}, which measures the amplitude reduction of the R-peak; and the ventricular residue (VR) similar to \cite{alcaraz2008adaptive}, which considers both the area and the amplitude of the QRS\footnote{QRS is the combination of three graphical deflections  (Q wave, R wave, and S wave)  on a typical electrocardiogram. } interval in the atrial activity.
	
	For an EGM containing $Q$ ventricular  segments, the  amplitude reduction of the R-peaks averaged over $K$ electrodes is 
	\begin{equation}
	\text{VDR} = \frac{1}{K}\sum_{i=1}^{K}\left(\frac{1}{Q}\sum_{q=1}^{Q}10\text{log}\left(\frac{R_{i,q}^{\text{m}}}{R'_{i,q}}\right)\right)
	\label{eq:vdr}
	\end{equation}
	where $R_{i,q}^{\text{m}}$ is the $q$th R-peak amplitude of the mixed EGM (in the time domain) of the $i$th electrode, and $R'_{i,q}$ is the amplitude of the respective residue. A higher value of VDR indicates more reduction of the ventricular activity.
	
	For an EGM containing $Q$ ventricular activity segments, the averaged VR is
	\begin{equation}
	\text{VR} =  \frac{1}{K}\sum_{i=1}^{K}\left(\frac{1}{Q}\sum_{q=1}^{Q}\left(\frac{ A_{i,q} \sqrt{\sum_{t=b_{i,q}}^{e_{i,q}}\left(a'_i(t)\right)^2}}{\sqrt{\frac{1}{T}\sum_{t=0}^{T}\left(a_i(t)\right)^2}}\right)\right)
	\label{eq:vr}
	\end{equation}
	where  $[b_{i,q}, e_{i,q}]$ is the $q$th QRS interval  in the estimated atrial activity of the $i$th electrode,
	and $A_{i,q}$ is the maximum amplitude in this interval. A lower value of VR indicates a better extracted atrial activity.

	{\textit{Results on synthetic data:}}
	For different degrees of atrial fibrillation, we evaluate the performace on the six segments and made the boxplots of results. Figure 11 compares the proposed GAE algorithm with the reference methods. The performance of the GAE algorithm [cf. (\ref{eq:extraction})] depends on the parameters $c$ and $\mu$.  These parameters are chosen based on a  grid search by minimizing the NMSE and are set to $c = 0.14$ and $\mu = 2$.
	We observe that the proposed method outperforms the other alternatives  by achieving the smallest NMSE and VR, and the largest CC and VDR for both degrees of atrial fibrillation. The ABS performs worse since it cannot adapt to changes in the EGM morphology caused by the heart activity variations. The performance of the AVC is unstable because it relies on the reference signal.
	The ICA performs poorly on this data since the independence assumption between the atrial and  ventricular activities  might not always hold in the EGM data.

	To further illustrate the differences of these methods, we show in Figure 12 an arbitrary  example of the synthetic EGM, the ground truth atrial activity, and the estimated atrial activity. We see that the  signal  extracted by the GAE method approximates the ground truth better than the reference algorithms. The ABS algorithm performs also well, but more of the  ventricular components is left compared to the GAE method.
	Also, the AVC and the ICA algorithms face difficulties in annihilating the ventricular component.

	{\textit{Results on real data}}:
	We move now on to the results  on the clinical EGMs. We evaluated the performance only through the non-intrusive metrics VDR [cf. (\ref{eq:vdr})] and VR [cf. (\ref{eq:vr})].  Table \ref{tb:performance_real} groups the results for the ten patients. For each patient, it reports the averaged performance over all electrodes and the respective standard deviation (in brackets). We see that the improved performance of the proposed GAE algorithm is further corroborated also with the real data.

	\begin{small}
		\begin{table}[]
			\centering
			\caption{Comparison of different algorithms for different patients during atrial fibrillation}
			
			\resizebox{0.49\textwidth}{!}{
				\setlength{\tabcolsep}{1mm}{
					\begin{tabular}{cccccc}
						\toprule	
						\multirow{2}{*}{Patient No.} & \multirow{2}{*}{Metrics} & \multirow{2}{*}{ABS} & \multirow{2}{*}{AVC} & \multirow{2}{*}{ICA} & \multirow{2}{*}{GAE} \\ 
						[0.3cm]\hline
						\multirow{2}{*}{P1}&VDR & 11.06 (3.31) & 7.99 (4.87)& 5.68 (5.39)  &$\mathbf{17.15}$ $\mathbf{(6.31)}$   \\ 
						&VR & 3.66  (2.08)& 7.86 (2.89) & 10.24 (1.84) & $\mathbf{1.47}$ $\mathbf{(0.50)}$  \\               
						\hline
						\multirow{2}{*}{P2}&VDR & 10.09 (3.43) & 7.96 (2.76)   & 6.37 (4.85)  &$\mathbf{16.98}$ $\mathbf{(4.40)}$ \\ 
						&VR & 2.93 (0.80) &  8.16 (1.66)  & 6.60 (2.58)  & $\mathbf{1.19 }$ $\mathbf{(0.31 )}$ \\ 
						\hline
						\multirow{2}{*}{P3}&VDR &  11.41 (4.26)   & 7.80 (3.67)  & 8.55 (4.42)  &$\mathbf{15.68 }$ $\mathbf{(4.34 )}$\\ 
						&VR & 3.17 (0.74)  & 7.08 (1.42)  & 6.71 (1.65)  &$\mathbf{1.64 } $ $\mathbf{(0.58 )} $\\ 
						\hline
						\multirow{2}{*}{P4}&VDR &  15.02 (4.12)  & 9.55 (4.27) & 7.42 (4.06)  &$\mathbf{16.85 }$ $\mathbf{( 3.43)}$\\ 
						&VR & 4.20 (0.50)  & 9.40 (2.21)  & 6.69 (1.35) &$\mathbf{1.80 } $ $\mathbf{(0.46 )} $\\ 
						\hline
						\multirow{2}{*}{P5}&VDR & 7.80 (3.26)    &  8.73 (4.76) & 6.59 (4.26)  &$\mathbf{ 14.51}$ $\mathbf{(3.57 )}$\\ 
						&VR &5.07 (0.67)  &8.94 (3.28)   & 10.20 (1.98)  &$\mathbf{2.65 } $ $\mathbf{(0.46)} $\\ 
						\hline
						\multirow{2}{*}{P6}&VDR &9.84 (2.97)      &8.39 (4.20)   & 5.84 (1.67) &$\mathbf{ 16.57}$ $\mathbf{(4.39 )}$\\ 
						&VR & 6.74 (1.14)  & 12.37 (2.88)  & 7.30 (1.67)  &$\mathbf{ 2.43} $ $\mathbf{( 0.53)} $\\ 
						\hline
						\multirow{2}{*}{P7}&VDR & 10.39 (4.21)     &6.86 (5.33) &  4.34 (1.63)  &$\mathbf{12.18 }$ $\mathbf{(4.64 )}$\\ 
						&VR & 3.03 (0.79)  & 9.43 (1.99)  &12.44 (2.16)  &$\mathbf{2.39} $ $\mathbf{(0.68)} $\\ 
						\hline
						\multirow{2}{*}{P8}&VDR & 5.72 (3.91)    & 5.27 (3.55)  & 5.94 (2.46) &$\mathbf{ 11.95}$ $\mathbf{(2.94 )}$\\ 
						&VR &4.36 (0.57)   & 8.59 (1.81)  & 12.40 (1.53)  &$\mathbf{ 2.60} $ $\mathbf{(0.66 )} $\\ 
						\hline
						\multirow{2}{*}{P9}&VDR &14.59  (4.62)   & 7.71 (4.35)  & 4.53 (3.79)  &$\mathbf{ 17.13}$ $\mathbf{( 5.54)}$\\ 
						&VR & 2.18 (0.74)  &12.70 (1.76)  & 13.21 (4.53) &$\mathbf{ 2.62} $ $\mathbf{( 0.76)} $\\ 
						\hline
						\multirow{2}{*}{P10}&VDR & 9.52 (4.57)    & 8.69 (5.05)  &  8.45 (4.25)  &$\mathbf{14.93}$ $\mathbf{( 5.01)}$\\ 
						&VR & 5.49 (0.74) & 8.83 (3.33) & 6.18 (2.09)  &$\mathbf{2.14 } $ $\mathbf{(0.94 )} $\\ 
						\hline
						\multirow{2}{*}{Mean}&VDR &10.55 (4.85)  & 7.90  (5.01)  &6.30 (4.26)  &$\mathbf{15.39}$ $\mathbf{( 4.92)} $\\ 
						&VR &4.08 (1.63)& 9.34 (2.54)  & 9.20 (1.12) &$\mathbf{2.09 } $ $\mathbf{( 0.79)} $\\ 
						
						\bottomrule

				\end{tabular}}
			}
			\label{tb:performance_real}
			\vspace{-0.4cm}
		\end{table}
	\end{small}
	Figure 13 shows a random example of the measured EGM and the  extracted atrial activity by the different algorithms.  The proposed GAE method extracts a smoother signal and has less ventricular component left.  The extracted signal by ABS presents more fluctuations   since ABS uses a fixed template to subtract the ventricular activity. The AVC shows a slightly better result  than ABS, but it has more ventricular components left. The ICA can remove the ventricular activity well but fails in preserving the atrial activity.

	\section{Discussion and Future Recommendations}
	
	We proposed an approach based on graph signal processing to analyze atrial fibrillation. This method combines the graph Fourier transform with the short-time Fourier transform to analyze multi-electrode epicardial electrograms in a joint space, time, and frequency domain. By working with a higher-level model, we tackled the difficulties of analyzing the disease through complicated physical models. We found a strong link between the spatial and temporal variation of the atrial signal and the atrial fibrillation leads to a reduction of signal spatial variation. We also characterized the space-time-frequency differences of the atrial and ventricular activities and developed a graph-based algorithm to estimate the atrial signal from the mixed measurements. The proposed algorithm corroborates our theory by showcasing improved performance with respect to other state-of-the-art methods.

	The proposed framework has also limitations. An initial difficulty we faced is how to construct the most representative graph. While we relied on a Euclidean-based nearest neighbor approach, it remains still an open question whether it is possible to find a more meaningful structure through graph learning techniques \cite{mateos2019connecting}. The used type of graph is, in fact, crucial since it gives the Fourier basis to capture the spatial variability. We believe that the performance of the graph-based extraction algorithm can be substantially improved if smooth-based graphs are learned \cite{kalofolias2016learn,qiu2017time}. Among the same lines, it remains unanswered whether directed graphs and other graph representation matrices (e.g., normalized or random walk Laplacian) can yield different insights on atrial fibrillation.

	It did not escape our notice that the graph-based extraction algorithm imposes a tradeoff between the preservation of the atrial activity and the reduction of the ventricular activity. The latter is heavily influenced by the smoothness upper-bound in (19). This parameter along with the Lagrange penalty term has been selected using a grid search. However, it deserves further investigation to check if constant values for different patients are a good choice or if we need to  adjust the values for each separate case. We also believe that other graph- and graph-time priors such as diffusion or bandlimitedness can impose a better tradeoff for atrial activity extraction \cite{marques2017stationary, perraudin2017towards}.

	Another direction worth taking in the near future is to corroborate our findings on a larger dataset, with induced and spontaneous atrial fibrillation, and to characterize the graph-time spectral behavior of the disease levels. In this direction, we also aim to adopt graph-based techniques to detect atrial fibrillation triggers from electrogram measurements.
	
	Altogether, our aim is to raise attention to explore spatial-temporal spectral properties of electrocardiograms to move forward the research of atrial fibrillation.

	\section{Conclusions}
	We suggested a new approach to study the  epicardial electrograms  for atrial fibrillation. This approach relies on graph signal processing--a recent research area in  the signal processing community--to model electrograms during atrial fibrillation with a higher level model.
	We conducted a novel graph-time spectral analysis study to analyze the epicardial electrograms in the joint space, time, and frequency domains. We found that the spatial variation
	is related to the high temporal variation; precisely, a faster temporal variation induces a high spatial variation. We  also found that the atrial fibrillation reduces the high temporal frequencies of the atrial electrogram. Together, these observations suggest that atrial fibrillation leads to a decrease of the spatial variation of the atrial activity. We also observed that the ventricular activity is smoother over the graph compared with the atrial activity. In this respect, we designed a graph-based atrial activity extraction algorithm that leverages the smoothness prior to estimate the atrial activity. Our experimental results with  synthetic data and real electrocardiograms showed that 
	the propose method outperforms reference methods that are based on average beat subtraction, adaptive ventricular cancellation and independent component analysis. These findings  shed light to new ways to approach the disease and  maybe of help to further understand its mechanisms.

	\bibliographystyle{elsarticle-num}
	\bibliography{cas-refs}


\end{document}